\documentclass[fleqn,usenatbib]{mnras}

\usepackage{newtxtext,newtxmath}
\usepackage{longtable}
\usepackage{lscape}
\usepackage{rotating}

\usepackage[T1]{fontenc}

\DeclareRobustCommand{\VAN}[3]{#2}
\let\VANthebibliography\thebibliography
\def\thebibliography{\DeclareRobustCommand{\VAN}[3]{##3}\VANthebibliography}

\usepackage{graphicx}	
\usepackage{amsmath}	

\usepackage{longtable}

\newcommand{\teff}{\ensuremath{T_\mathrm{eff}}}
\newcommand{\logg}{\ensuremath{\log\,g}}

\newcommand{\Vsini}{$V \sin{i}$}
\newcommand{\kms}{km s$^{-1}$}

   \title[Rotation of B stars in Carina]{The Gaia-ESO Survey: Projected Rotational Velocities of B stars in the Carina Nebula
   \thanks{Based on observations collected at the ESO telescopes under programme 188.B3002, 193.B-0936, and 197.B-1074, the Gaia-ESO Public Spectroscopic Survey.}
   }

\author[W. Santos et al.]{
W. Santos$^{1,2}$,
S. Daflon$^{1}$,
J. V. Sales Silva$^{1}$,
K. Cunha $^{1,3}$, 
R. Blomme$^{4}$, 
T. Morel$^{5}$, 
A. Herrero$^{6,7}$, 
\newauthor
J. Maiz Apellániz$^{8}$,
L. Mahy$^{4,5,9}$,
S. R. Berlanas$^{6,7}$,
T. Bensby$^{10}$,
A. Bragaglia$^{11}$,
F. Damiani$^{12}$,
A. Frasca$^{13}$,
\newauthor
G. Gilmore$^{14,15}$,
V. Kalari$^{16}$,
A. Lanzafame$^{13}$,
S. Randich$^{17}$,
S. Van Eck$^{17}$,
S. Zaggia$^{18}$
L. O. Kerber$^{2}$
\newauthor
O. J Katime Santrich$^{2}$
\\
$^{1}$Observat\'orio Nacional (ON)-MCTI, Rua Jos\'e Cristino, 77, 20921-400 Rio de Janeiro, RJ, Brazil
  \thanks{Contact e-mail: wjsantos@uesc.br}
\\
$^{2}$Universidade Estadual de Santa Cruz - UESC, Rodovia Ilh\'eus/Itabuna, km 16 - 45662-900 Ilh\'eus, BA, Brazil
\\
$^{3}$ Steward Observatory, University of Arizona, Tucson, AZ 85721, USA
\\
$^{4}$Royal Observatory of Belgium, Ringlaan 3, B-1180 Brussels, Belgium 
\\
$^{5}$Space Sciences, Technologies, and Astrophysics Research (STAR) Institute, Universit\'e de Li\`ege, Quartier Agora, B\^at B5c, Allee du 6 ao\^ut, 19c, 4000 Lie\`ege, Belgium
\\
$^{6}$Instituto de Astrofisica de Canarias, E 38200 La Laguna, Tenerife, Spain
\\
$^{7}$Departamento de Astrofisica, Universidad de La Laguna, E-38206 La Laguna, Tenerife, Spain
\\
$^{8}$Centro de Astrobiologia (CSIC-INTA), ESAC campus. Camino bajo del castillo s/n, E-28\,692 Villanueva de la Ca\~nada, Madrid, Spain
\\
$^{9}$Instituut voor Sterrenkunde, KU Leuven, Celestijnlaan 200D, Bus 2401, 3001 Leuven, Belgium
\\
$^{10}$Lund Observatory, Division of Astrophysics, Department of Physics, Lund University, Box 118, SE-22100 Lund, Sweden 
\\
$^{11}$INAF – Osservatorio di Astrofisica e Scienza dello Spazio di Bologna, via P. Gobetti 93/3, 40129, Bologna, Italy
\\
$^{12}$INAF – Osservatorio Astronomico di Palermo, Piazza del Parlamento 1, I-90134, Palermo, Italy
\\ 
$^{13}$INAF – Osservatorio Astrofisico di Catania, Via S. Sofia 78, 95123 Catania, Italy
\\
$^{14}$Institute of Astronomy, University of Cambridge, Madingley Road, Cambridge CB3 0HA, United Kingdom
\\
$^{15}$Institute of Astrophysics, FORTH Cert, N. Plastira 100, GR-70013 Vassilika Vouton, Crete, Greece
\\
$^{16}$Gemini Observatory/NSF's NOIRLab, Casilla 603, La Serena, Chile
\\
$^{17}$INAF – Osservatorio Astrofisico di Arcetri, Largo E. Fermi 5, 50125, Firenze, Italy
\\
$^{18}$INAF – Osservatorio Astronomico di Padova, Vicolo dell’Osservatorio 5, 35122, Padova, Italy
}

\date{Accepted XXX. Received YYY; in original form ZZZ}

\pubyear{2023}

\begin{document}
\label{firstpage}
\pagerange{\pageref{firstpage}--\pageref{lastpage}}

\maketitle

\begin{abstract} 
   The Carina Nebula is an active star-forming region with several open clusters rich in massive OB stars, thus making it an optimal target for studying stellar properties such as rotation for large samples of these early-type stars. We  studied a sample of early-type stars probable members of the 8 open clusters in the Carina Complex. The observational data consist of high-resolution spectra from the Gaia-ESO public Spectroscopic Survey. Astrometric and photometric data from Gaia EDR3 and radial velocities measured from the observed spectra are used to confirm the cluster members. The projected rotational velocities of 330 early-type stars of Carina are derived from the widths of \ion{He}{i} lines at 4388 and 4471 \AA. The reported  \Vsini\ values are the first estimates for 222 early-type stars.  The \Vsini\ distribution for the Carina clusters peaks at $\sim$100-150 \kms, consistent with the distributions for B stars in Galactic clusters.  \Vsini\ estimates for stars members of the clusters  Trumpler 15, Collinder 228, Collinder 232, and Bochum 11 are presented for the first time in the literature.  For a subsample of stars with earlier spectral types from B0 to B3, we find a bimodal distribution, with a third, small peak towards the upper values of \Vsini. When the full sample is split according to the parent cluster, we find that the oldest cluster in our sample, NGC 3293, presents a higher concentration of rapidly rotating stars. In contrast, Collinder 228 presents a larger number of stars with lower \Vsini.
\end{abstract}
  
%




\section{Introduction}

The Gaia-ESO Survey \citep[GES;][]{Gilmore2012,Gilmore2022, Randich2022} performed one of the largest high-resolution spectroscopic surveys of the Galaxy, aiming to characterize the kinematics and abundances of the Milky Way through a sample of $~ 10^{5}$ stars observed in the different components of the Galaxy (Bulge, Thin and Thick Disks and Halo).
This extraordinary sample includes the Carina Nebula,
which is a region with a high concentration of gas and dust,  providing an ideal interstellar medium to trigger and maintain star formation processes, as shown by \citet{Rebolledo2020}.
The  Car OB1 association is a large,  very active star-forming region in the  Carina-Sagittarius spiral arm,  and its stellar population is split into several open clusters \citep{Lim}.
The Carina Nebula and its associated clusters have been studied in detail in a large number of photometric studies, such as those by \citet{feinstein1973, Cudworth1993, Massey1993} and \citet{ hur2012}.

Carina has been the target of many spectroscopic studies, including the analysis of cool stars by \citet{Damiani2017}.
Based on X-ray data, \citet{Damiani2017}
 investigated new possible low-mass candidate members in Carina obscured by the dark nebula.
The nonuniform reddening in Carina is remarkable due to the presence of significant dust lanes throughout the nebula: the inner region presents a larger ratio of $E(V-I)/E(B-V)$ in comparison with the outskirts.
More recently, \citet{jesus2018} determined that the extinction across the Carina Nebula is variable not only on the amount of extinction [$E(4405-5495)$] but also on its type [$R_{5495}$], revealing the existence of different environments along the line of sight of different stars. This issue had previously caused significant discrepancies in the historical measurements of the distance to the Carina nebula (see \citet{VillafrancaI} for a review). The ISM in Carina has already been discussed by \citet{WalbornHesser} and \citet{Walborn2002}.

The Villafranca project \citep{VillafrancaI, VillafrancaII} is producing a census of Galactic stellar clusters with OB stars, including some of the clusters and subassociations in the Car~OB1 association. More recently, a third paper of the project (Molina Lera et al. in preparation) has analyzed the whole Car~OB1 association. In those papers, it was found that Trumpler~14 and Trumpler~15 are likely real (i.e. bound) clusters and that Trumpler~16~ can be divided into two subregions (Trumpler~16~W and Trumpler~16~~E) that are separating from each other as a result of an overall expansion. On the other hand, Collinder~228, Collinder~232, and Bochum~11 are not clearly defined clusters but instead are regions (subassociations) of the Car~OB1 association. Other parts of the association are located farther away. One of them,   NGC~3293, is an independent massive cluster somewhat older than the previously mentioned ones.

 Stellar rotation is fundamental to understanding the formation and evolution of early-type stars, considering that the distribution of rotational velocities as a function of spectral types reaches its maximum  
at early-B stars  \citep{stauffer1986}. Moreover, models of stellar evolution including rotation show that this parameter alters the duration of the main sequence and induces mixing processes \citep{Evans2005}. 
Some spectroscopic studies have analyzed the physical parameters of stars in the Carina Nebula, including Projected Rotational Velocities (\Vsini).
Estimates of stellar parameters and rotational velocities based on spectroscopic analysis are 
available in the literature for stars in Trumpler 14 and Trumpler 15 \citep{Hanes2018}, Trumpler~16~ and NGC 3293 \citep{Huang2006, Huang2006a}. NGC 3293 is part of the sample studied by  \citet{Wolff2007} to investigate the relation between stellar rotation and environment. More recently, based on the Fourier transform, \citet{berlanas23}  obtained rotational velocities for a sample of O stars in Carina. 
Nevertheless,  approximately
two-thirds of the sample studied in this sample still lack rotational estimates, which are provided for the first time in this paper. 


In this paper, we present 
measurements of radial velocities and estimates of \Vsini\
derived from the widths of  \ion{He}{i} lines for a sample of main-sequence early-type stars. We also analyze for the first time the \Vsini\ distributions of the open clusters in central region of the Carina Complex.
This paper is organized as follows:  
in Sect. ~\ref{data} we describe the observational data; in Sect. ~\ref{member} we present a short description of the membership analysis and the adopted criteria for target selection; in Sect. ~\ref{vsini}, we explain the methodology used to obtain the \Vsini\ values.  In Sect.~\ref{Discussion}, we discuss our \Vsini\ results in terms of the cluster distributions and comparison with previous results from the literature. In Sect. ~\ref{Conclusions}, we summarize our conclusions.


\section{Data}\label{data}

This paper presents the results of a spectroscopic analysis for a sample of early-type stars belonging to open clusters in the Carina region. The study has been conducted by one of the working groups of the GES collaboration, located at the Observatório Nacional, Brazil (so-called ON Node), 
based on its sixth Data Release (DR6). 
The sample of stars was defined from a photometric analysis based on color-magnitude diagrams to select targets with a high probability of being cluster members, as described in  \citet{Blomme} and \citet{Bragaglia}.
 
\begin{figure}
   \centering
   \includegraphics[width=\hsize]{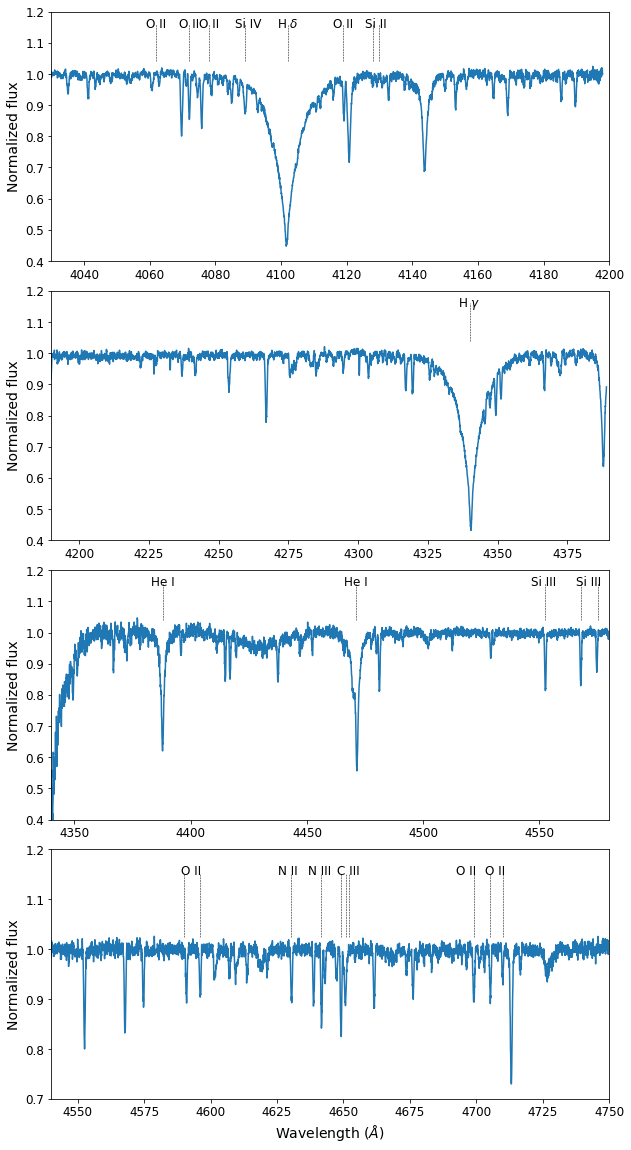}
      \caption{Examples of observed spectra for the B1 star ALS~15861, corresponding to the four GIRAFFE configurations HR03, HR04, HR05A, and HR06. Relevant absorption features are identified in the panels. }
         \label{spec}
   \end{figure}

The spectra analyzed in this paper have been obtained with the GIRAFFE multifiber spectrograph coupled to the UT2 of the Very Large Telescope (VLT) at ESO, Chile. 
The GIRAFFE setups used by GES are optimized to detect important features in the spectra of early-type stars: HR03, with spectral coverage $\lambda\lambda$ 4033 - 4201\AA, contains H$\delta$ and \ion{Si}{ii} lines; HR04, which covers from $\lambda\lambda$ 4188 to 4392 \AA, contains the H$\gamma$ line, an important \logg\ indicator; HR05A has \ion{He}{i} and \ion{Si}{iii} lines and runs from $\lambda\lambda$ 4340 to 4587\AA; and  HR06 covers $\lambda\lambda$ 4538 - 4759 \AA\ and presents some important lines for \teff\ diagnosis, such as    \ion{C}{iii}, \ion{O}{ii}, \ion{Si}{iii}, and \ion{Si}{iv} lines. 
The observed spectra were normalized by the ROBGrid node \citep{Blomme}  and have a nominal resolution typically higher than 20,000 \citep{Gilmore2022,Randich2022}. 
Examples of the observed spectra of the B-type star ALS~15861 are shown in the four panels of Figure~\ref{spec}.
The spectral ranges correspond to the four instrumental configurations HR03, HR04, HR05A, and HR06, from top to bottom. Relevant absorption features are identified in the panels.

\begin{figure}
\centering
\includegraphics[width=\hsize]{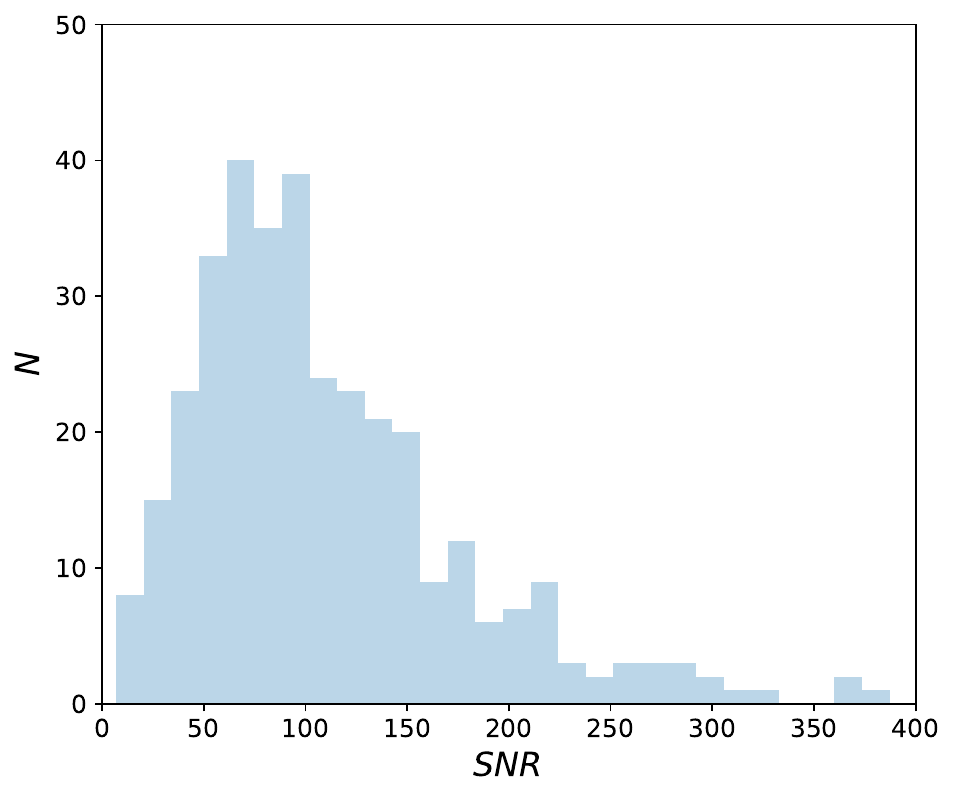}
\caption{Distribution of Signal-to-noise ratios measured in spectra from the sample of 347 stars of spectral types B5 to O9.5 obtained through the HR5 setup.}
\label{S_N}
\end{figure}

 The signal-to-noise ratio (SNR) analysis provides a valuable assessment of the spectrum quality and the reliability of subsequent studies.
The SNR measurement of the observed spectra was conducted in the continuum region around $\lambda$4455 \AA, using the IRAF task {\it splot}. The measured SNR values are listed in column 4 of table~\ref{tab:vsini} and
the histogram of the SNR results is shown in Figure~\ref{S_N}.  Most of the analyzed spectra have an SNR of at least $\sim 50$, with an average of 112.

The GES obtained GIRAFFE spectra with the blue gratings for 2112 stars in the direction of the Carina Nebula, providing a sample covering different spectral types and luminosity classes.
The entire extent of Carina has not been covered by GES observations due to the observational limitation of the survey \citep{Bragaglia}. 
Since our analysis focuses on early-type stars, we used the preliminary temperatures provided by GES as an initial criterion to select stars with 14,000 $<$\teff (K) $<$33,000 to form our sample.
Stars with spectra showing asymmetric or duplicated lines (indicative of double-lined spectroscopic binary stars) and strong emission profiles in the Hydrogen lines (evidence for strong winds or disks) \footnote{Stars with at least one Hydrogen line free of emission in their spectra were kept in the sample.} have been discarded from our sample, since these features may introduce additional constraints not included in the methodology adopted in this study. 
Potential single-lined spectroscopic binaries (SB1) in our sample were not flagged, as the GES observations were not specifically designed for binary analysis. \citet{Morel} conducted a detailed examination of the spectra of stars in NGC 3293, obtained with different instrumental configurations and at different epochs. Their study reported a lower-than-expected binary fraction for typical early-type star samples, likely due to the survey design. Consequently, in our analysis, SB1 candidates and single stars were treated using the same methodology.
Furthermore, a few stars exhibiting either extremely weak or absent diagnostic lines—due to very low signal-to-noise ratio (SNR) or exceptionally high \Vsini—were excluded from the sample. As a result, our analysis begins with a selection of 347 single-lined, early-type stars associated with open clusters in the Carina Nebula.

\section{Membership}\label{member}

The Carina nebula is a very complex region due to its numerous nebular and stellar components, including the clusters Trumpler 14, Trumpler 15, Trumpler~16~E, Trumpler~16~W, Collinder 228 Collinder 232, Bochum 11, and NGC 3293. The gas in this region exhibits an intense temperature variation, ranging from cold  \citep{Yonekura} to hot  \citep{Seward,Townsley}.
Carina remains a highly active star-forming region, featuring dense pillars of dust and bright emission from \ion{H}{ii} regions \citep{Povich}. The clusters are embedded in molecular clouds that produce significant extinction, with $A_{v}$ $\sim$ 15 mag \citep{Damiani2016}.  
A CO-rich molecular cloud with V-shaped dust lanes can be seen in the center of the OB association Car OB1 \citep{Rebolledo2016}.  This dust lane likely plays an important role in the differential extinction observed in the Carina Nebula \citep{jesus2018}.  
A recent census of OB stars in the Carina Nebula
 \citep{berlanas23} discussed the completeness in the region, considering the high extinction observed. Their associated catalog lists 316 OB stars in Carina, representing an almost threefold increase in the number of previously known stars.

 Since the targets of interest are located very close to the Galactic plane (b$\sim$ 0.06$^o$ -- 0.07$^o$) and the fraction of contaminants may be significant     \citep[$\sim$ $20\%$;][]{Bragaglia}, we performed an additional analysis of the astrometric and kinematic properties of targets observed by the Gaia mission and radial velocities estimated by us through GES spectra.

 To determine the stars of each open cluster in Carina, 
 we used the same procedure for membership analysis as applied by other GES groups for cooler stars.
 Astrometric and photometric data from Gaia EDR3 \citep{Gaia_Collaboration_Brown_2021} were used to define the members of each cluster.
We selected stars
from the Gaia EDR3 catalog \citep{gaia-edr3} 
within a radius of 10' centered on the studied clusters and cross-matched them with our list of stars observed with GES. We used the Renormalised Unit Weight Error (RUWE)  measurement to discard stars with poor astrometric solutions. The RUWE value represents the goodness of fit, with  < 1.0 indicating a good astrometric source. RUWE values between 1.0 and 1.4 may indicate an astrometric binary, a (partially) resolved binary or multiple stars, or even a problematic source \citep{Lindegren2020}.
Approximately 62$\%$ of the stars in our sample have RUWE < 1.0; for $\sim$ $27\%$  of the stars in our sample, RUWE measurements are within the range 1.0 -- 1.4. Finally, for less than  $\sim$ $11\%$ of our sample, RUWE is larger than 1.4, which may indicate problems with the processing of astrometric measurements or a poorly behaved solution. Stars with RUWE > 1.4 have been discarded from our analysis. As an additional parameter for a good astrometric solution from Gaia, we also considered precise parallax ($\varpi$) measurements, with 
$\sigma_{\varpi} \sim 10\%$.

 The parallax $\varpi$ has been corrected by the zero point ZEDR3, following \citet{Lindegren2021},  to obtain the corrected parallax, $\varpi_{c}$. We followed the procedure described in \citet{Jesus2021}  and \citet{VillafrancaII} to correct the internal parallax, in case of possible underestimates. 
The corrected parallaxes are slightly larger than the original results, representing small variations  $\delta \varpi$ $<$0.03 $mas$ for the clusters studied in this paper. In general, these corrections produce small offsets in the parallax results, leading to more accurate distances.

\begin{figure}
\centering
\includegraphics[width=\hsize]{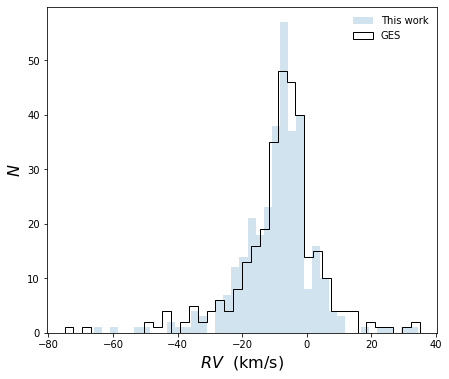}
\caption{Comparison between the distributions of radial velocities obtained in this work (light blue histogram) and the values from other GES nodes (black-lined histogram).}
\label{rv}
\end{figure}

The radial velocities were determined through cross-correlations between synthetic spectra at rest reference and the observed spectra of the target stars. This task was accomplished using the absorption lines of \ion{He}{i} at  $\lambda\lambda$4388, and 4471 \AA\  and \ion{Si}{iii}  at  $\lambda\lambda$4552, 4567, and 4574 \AA\ present in the HR05A spectra.
The measured radial velocities are listed in Table~\ref{tab:vsini} and Table~\ref{tab:nomember}.
 The distribution of the RV results for our sample stars is shown in Figure ~\ref{rv} as a blue histogram, in comparison with the distribution of RV measurements obtained by other GES nodes \citep{Hourihane23} for Carina (black-lined histogram). The two distributions are consistent, with the distribution of RVs provided by GES  presenting a slightly higher dispersion.

\begin{figure*}[t!]
   \centering
   \includegraphics[width=\hsize]{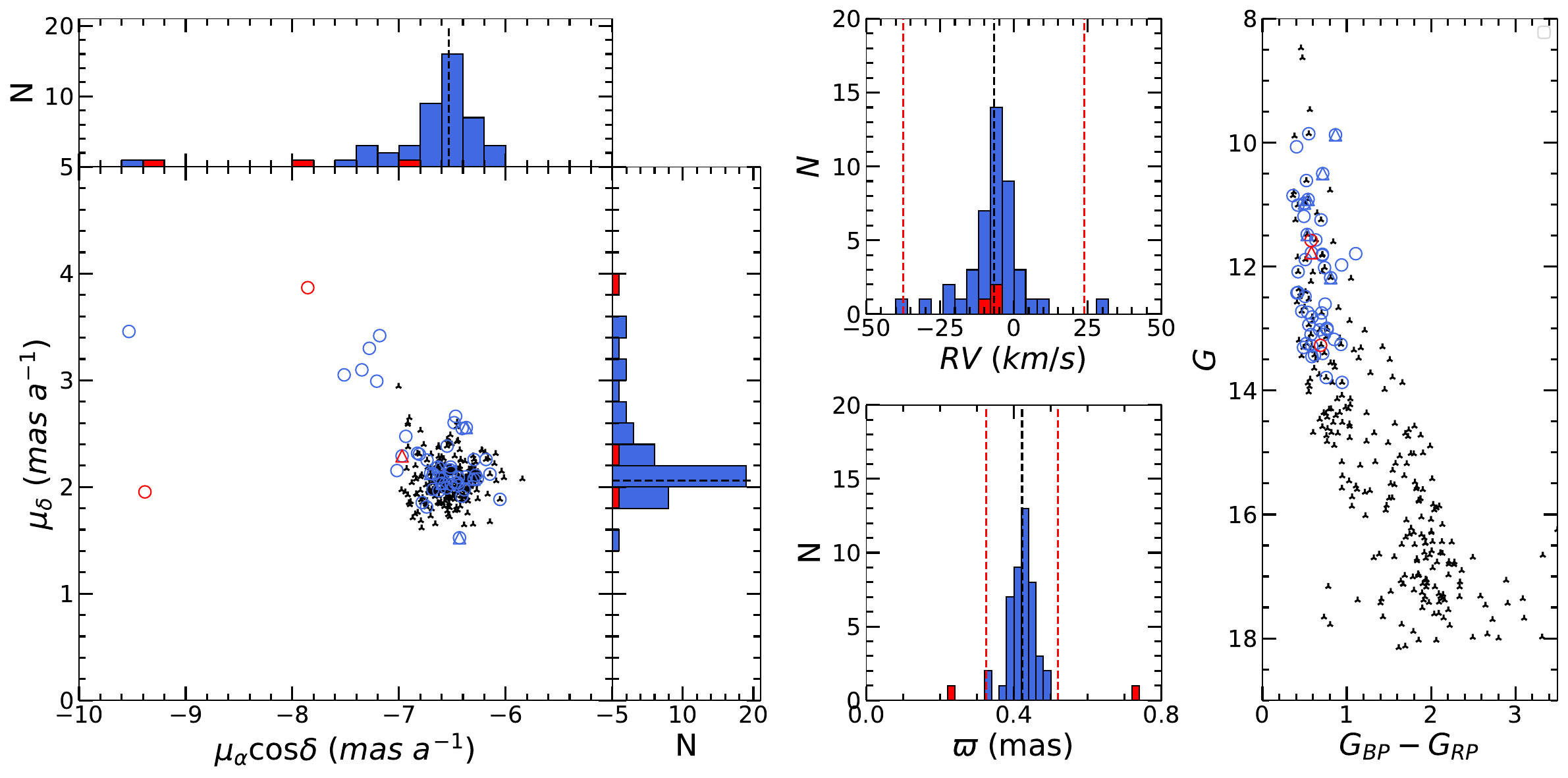}
      \caption{Gaia EDR3 astrometric and photometric data for the cluster Trumpler 14. Blue symbols represent stars that are cluster members, while red symbols highlight non-members. The open circles represent stars with \Vsini \ measurements, while those stars selected for abundance analysis are represented by open triangles. 
      The black small triangles represent the other cluster members by \citet{cantat}. {\it Left panel}: the distribution of proper motions in right ascension and declination. The side histograms are color-coded according to the symbols and the dashed line depicts the mean value for our sample; {\it Upper middle panel}: radial velocity distribution of the cluster stars; {\it Lower middle panel}: the distribution of parallax color-coded for probable members (blue) and probable non-members (red). The black dashed line is the mean value for our sample while the red dashed lines represent the 3$\sigma$ limits; and {\it Right panel}: the color-magnitude diagram for the cluster, using Gaia magnitudes.}
         \label{member_tr14}
   \end{figure*}

 Figure ~\ref{member_tr14} 
 presents the astrometric and photometric data provided by Gaia EDR3 for stars in the cluster Trumpler 14. 
 The left panel shows the distribution of proper motions in right ascension and declination.  
The black small triangles represent the cluster members listed by \citet{cantat}. Blue open circles represent cluster members while red open circles depict non-members. 
The open triangles with the same color code refer to the stars selected for the abundance analysis, which will be presented in a forthcoming paper. 
The side histograms are color-coded according to the open circles and the dashed line depicts the mean value for our sample. The upper middle panel shows the 
distribution of measured radial velocities whereas the distribution of parallax color-coded for probable members (blue) and probable non-members (red) is shown in the lower middle panel. In these two histograms, the black dashed lines are the mean value for the sample while the red dashed lines represent the 3$\sigma$ limits. Finally, in the right panel, we present the color-magnitude diagram  $({G_{RP}-G_{BP}}) \times G$. As expected, the selected sample of early-type stars occupies the upper left part of the CMD.
Figures from \ref{member_tr15} to \ref{member_ngc3293} are presented in the Appendix ~\ref{cmds}, for the clusters 
Trumpler 15, Trumpler~16~E, Trumpler~16~W, Bochum 11, Collinder 228 Collinder 232,  and NGC3293, respectively, and using the same symbols and colors as in Figure  \ref{member_tr14}.  The CMDs shown in  Figure ~\ref{member_tr14}  and Appendix ~\ref{cmds} are not corrected for extinction.

To define our final sample of stars in each cluster, we adopted a conservative selection criterion based on a 3$\sigma$ deviation in the parallax, similar to \citet{Goppl2022} and \citet{Shull2021}.
Stars with $\delta \varpi < 3\sigma$ or negative $\varpi$ are classified as background probable non-members,  while stars with  $\delta \varpi > 3\sigma$ are classified as foreground probable non-members.

The mean parallax for Trumpler 14 is 0.422$\pm$0.032 mas. Three stars deviate more than 3$\sigma$ from the cluster mean value; however one star is not shown in Figure~\ref{member_tr14} because its parallax value is negative and out of interval shown in the figure. The star ALS 15 864 has a parallax relatively close to the threshold value, and its radial velocity is consistent with the cluster mean value, so we decided to keep this star in the sample, marked as red open triangles in Figure ~\ref{member_tr14}. \citet{berlanas23} also marked this star as a probable member.  
ALS 15206 is reported in different sources in the literature either as a member of Trumpler 14 or Trumpler~16. Our analysis is more consistent with  ALS 15206 being a member of Trumpler 14,  in agreement with the conclusion of \citet{berlanas23}.

The mean parallax of Trumpler 15 is  0.425 $\pm$ 0.031 mas, and the parallaxes of six stars are outside the 3$\sigma$ limit. Two stars, ALS 15860 and GES 10451811-5924277, have parallax very close to the threshold value and were kept in the final sample for the chemical analysis. The rich cluster Trumpler~16~E has a mean parallax of 
0.430$\pm$0.023 mas, with two stars outside the 3$\sigma$ limit (one of them is not shown in the histograms of  Figure~\ref{member_tr16e} because its parallax value is off-scale). Trumpler~16~W has a mean parallax of 
0.430$\pm$0.027 mas, and the parallaxes of all stars fall well within the threshold limit.
NGC 3293 is the most populated cluster of our sample, with a mean parallax of 
0.420$\pm$0.031 mas. One star's parallax is outside the threshold, and it is considered a probable non-member. 

Bochum 11 has a mean parallax of 
0.432$\pm$0.028 mas, and one star, ALS 16082, may be considered a probable non-member, according to the 3$\sigma$ criterion. 
Collinder 228 is the cluster of our sample with the highest dispersion in parallax: 0.415$\pm$0.051 mas.   
Ten stars fall outside the 3$\sigma$ limit in parallax, although their radial velocities are consistent with the cluster distribution.  
Five of these stars, which are very close to the threshold, are good candidates for chemical analysis so we decided to keep them in the sample. The star 2MASS J10442909-5948207 was marked as a member of Collinder 228 while ALS 1853 was considered a background object by \citet{berlanas23}.   The mean parallax  of the cluster Collinder 232 is 
0.436$\pm$0.013 mas, and all stars are probable members according to the 3$\sigma$ criterion. 

Based on the analysis of the Figures \ref{member_tr14} and \ref{member_tr15} to \ref{member_ngc3293}, we found 23 background and foreground objects likely to be contaminants and those are listed in table~\ref{tab:nomember}. 
However, we decided to keep in our sample  8 of these stars that present parallax slightly outside the $3\sigma$ limit but have radial velocity and proper motion still consistent with the mean value of the parent cluster and are good candidates for the chemical analysis.  
At this stage, our sample comprises 332 stars that are probable members of the studied clusters: 
46 stars in Trumpler 14, 
40 stars in Trumpler 15, 
71 stars in Trumpler~16~E, 
 9 stars in Trumpler~16~W, 
19 stars in Collinder 232, 
22 stars in Bochum 11, 
59 stars in Collinder 228, and finally, 
66 stars in NGC 3293.

\section{Projected rotational velocities}\label{vsini}

\begin{figure*}[]
   \includegraphics[width=0.48\textwidth]{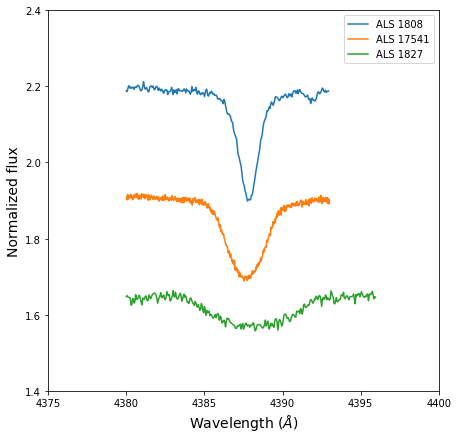}
   \includegraphics[width=0.48\textwidth]{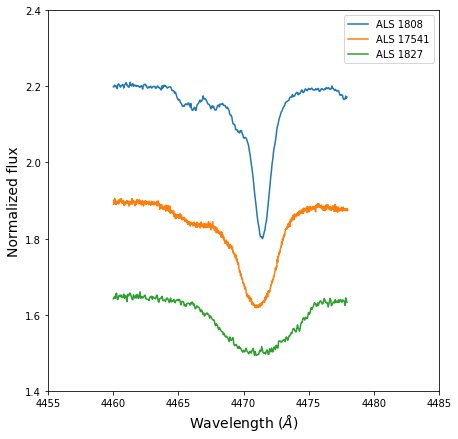}
       \caption{The effect of \Vsini\ on the observed profiles of \ion{He}{i} lines at 4388\AA\ (right panel) and 4471\AA\ (left panel). The \Vsini\ values corresponding to each example spectrum are, from top to bottom, 35 \kms (blue line,  star  ALS 1808),  114 \kms (orange line, ALS 17541), and 250 \kms (green line,  ALS 1827).}
    \label{linhas_4388_4471}
  \end{figure*}

The projected rotational velocity \Vsini\ becomes an important broadening mechanism at \Vsini\ greater than 100 \kms\ and significantly affects the moderate to weak metallic lines present in the spectra of early-type stars. 
However,  Helium lines can still be identified and measured in the observed spectra of rapidly rotating stars. The two panels of  Figure~\ref{linhas_4388_4471}
show the effect of \Vsini\ on the Helium lines at 4388\AA\ (right panel) and 4471\AA\ (left panel) of three stars with different rotational velocities: ALS 1808 with \Vsini=35 \kms, ALS 17541 with  \Vsini=114 \kms, and ALS 1827 with \Vsini=250 \kms. 

One of the aims of this analysis is to 
present \Vsini\ estimates for a sample of B stars in the carina Complex, based on the widths of Helium lines. 
We used the methodology described by \citet{Daflon2007}, based on the full width at half maximum (FWHM) measurements of synthetic profiles of three \ion{He}{i} lines at 4026, 4388, and 4471\AA, to estimate  \Vsini.
However,   our spectra cover only the lines  4388 and 4471\AA. Thus, the  \Vsini\ estimates are derived through the interpolation of the measured  FWHM of these two Helium lines in the grid of synthetic widths. This grid was built from measurements of the widths of non-LTE synthetic  \ion{He}{i} profiles computed with LTE model atmospheres for solar abundance \citep[ATLAS9;][]{Kurucz1993} combined with Non-LTE line formation  \citep[DETAIL;][]{giddings1980};  \citep[SURFACE;][]{butler1985}. The grid covers  \teff\ from 15,000 to 30,000 K, in steps of 5,000 K,  and constant values of surface gravity (\logg = 4.0, typical of young B stars) and microturbulent velocity (5 \kms). The synthetic spectra have been convolved with instrumental broadening corresponding to R $\sim$ 10,000 and 50,000, and rotational profiles ranging from \Vsini= 0 to 400 \kms,  in steps of 50 \kms. 
 The computations also included limb darkening using a 
linear limb-darkening law with a coefficient of $\sim$0.3, typical for early-type stars \citep{wade1985}.
\citet{Daflon2007} investigated the impact of surface gravity variations on \Vsini\  estimates and found that differences in \Vsini\ are negligible for changes in $\delta$(\logg) = $\pm$ 0.3 but can increase to 10–15\% for variations of $\delta$(\logg) = $\pm$ 0.5 dex. They concluded that, although the grid has been computed for a fixed value of \logg=4.0,  it can be safely used for stars with \logg = [3.7 -- 4.3].  

\begin{figure*}[h!]
   \includegraphics[width=0.48\textwidth]{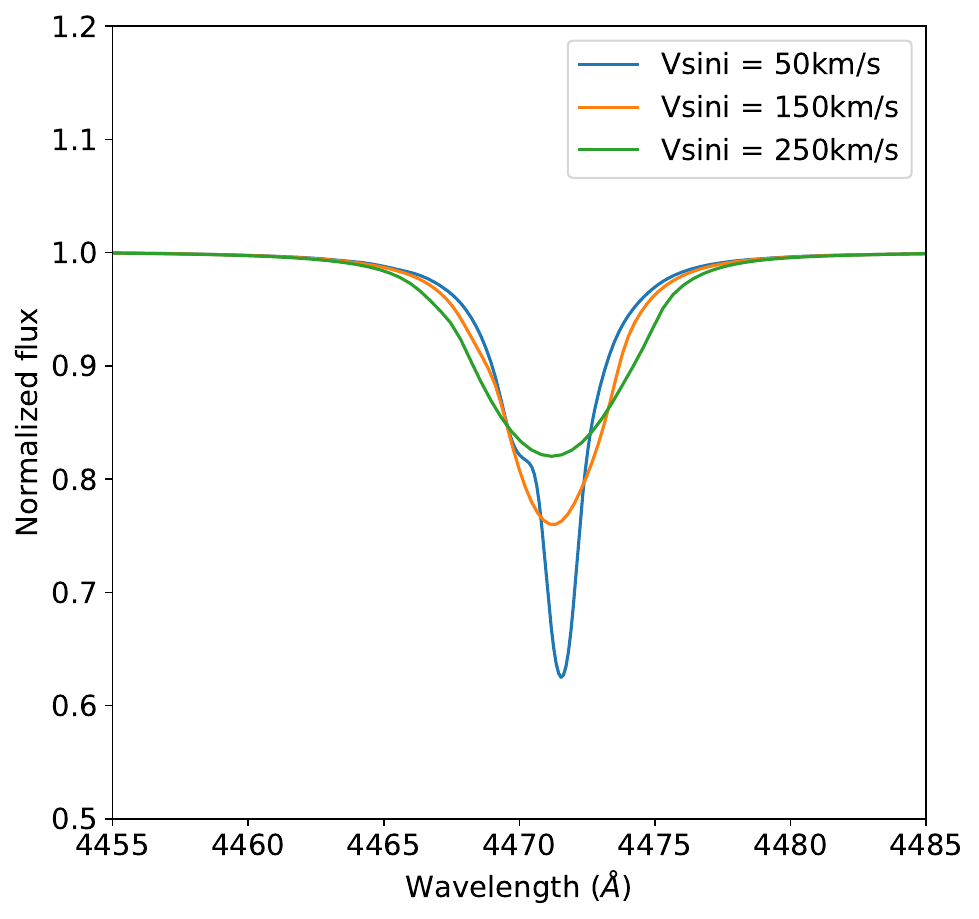}
   \includegraphics[width=0.48\textwidth]{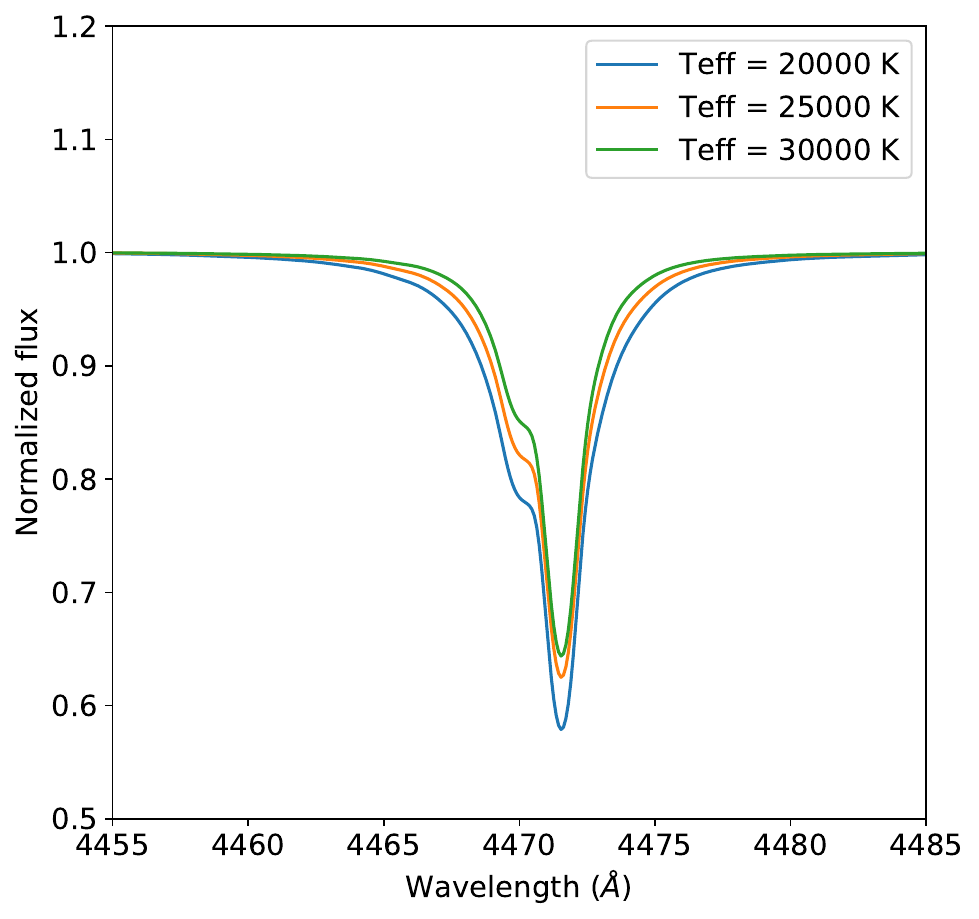}
       \caption{The effect of three different  \Vsini\ (left panel) and  of three different \teff\ (right panel) on the synthetic  profiles of \ion{He}{i} line 4471\AA.} 
    \label{synthetic}
  \end{figure*}

Upon examining the color-magnitude diagrams (CMDs) of the clusters in Figures \ref{member_tr14} to \ref{member_ngc3293}, we observe that most of the studied stars, indicated by blue circles or triangles, are either on the main sequence (MS) or have recently evolved from it. 
 To further investigate, we analyzed the observed hydrogen line profiles of six seemingly evolved stars—four from Trumpler 16E and one from Trumpler 16W. The width of the hydrogen line wings did not provide evidence of low surface gravity, suggesting that these stars may not be as evolved as their positions on the CMDs imply. This suggests that their surface gravities (\logg) likely fall within the expected range for such evolutionary stages.
 Finally, two stars in Trumpler 15 and four in Collinder 228 that appear to have evolved away from the MS and are marked with red symbols, have been classified as non-members.

\citet{Daflon2007} did not account for gravitational darkening and macroturbulence in their calculations, which can lead to underestimations of \Vsini\ by 12\% to 33\%, depending on the spectral type and the rotational regime. As a sanity check,  they also derived \Vsini\ values from the synthesis of weak metal lines for a subset of sharp-lined stars in their sample, finding excellent agreement with those obtained from the widths of \ion{He}{i} lines. The mean difference between the two methods was $-0.4 \pm7.7$ \kms, suggesting that this method can be safely used to obtain first estimates of rotational velocities.

 Following the recommendation of \citet{Daflon2007}, we considered that the grid is adequate for our sample since most of the stars in our sample are on the main sequence.
 The input parameters necessary to interpolate in the grid and obtain \Vsini\ are the effective temperature and the line widths measured from the observed spectra. 
 The widths of the  \ion{He}{i} lines 4388 and 4471\AA\  were measured using the task splot of the reduction package IRAF.
 This tool allows the manual fit of a Gaussian or Voigt profile to spectral lines and automatically computes the Full Width at Half Maximum (FWHM) from the fit parameters. The quality of the fit is influenced by the spectral signal-to-noise ratio (S/N), line depth, and blending with neighboring features. In our analysis, we adopted Gaussian fits for the \ion{He}{i} lines, and we verified consistency through repeated measurements. Based on the typical S/N ($>$50) in our spectra and comparisons between multiple lines per star, we estimate that the uncertainty in the FWHM measurements is in the range of 5–10\%. For spectra with S/N below $\sim$40, the uncertainties may increase significantly due to continuum placement errors and poor line contrast.

At this point of our analysis, the adopted effective temperatures  are those values recommended by GES. This approach is justified since the  \teff\ value has a small impact on the derived \Vsini. For example, considering the line \ion{He}{i} 4388\AA\ with FWHM=3.0\AA\, the obtained \Vsini\ is 107 \kms\ for  \teff=20,000~K and  105 \kms\ for  \teff=15,000~K.  
 Figure \ref{synthetic} illustrates the effects of \Vsini\ (left panel) and \teff\ (right panel) on the synthetic profiles of the line \ion{He}{i} 4471\AA.

 The estimation of \Vsini\ was not possible for those stars of our sample with very broad or very sharp lines, implying that  \Vsini\ is likely higher than 400 \kms or close to zero, respectively, the limits of the adopted grid. 
 These constraints led us to discard only 2 stars from the sample: 2MASS J10461906-5957543, with \Vsini\ $\sim$ 0, 
 and ALS 15248, with \Vsini\ $>$  400 \kms.
 Our final sample comprises 330 stars probable members with \Vsini\ estimates.
 In Table~\ref{tab:vsini} we present 
 the FWHM of the two \ion{He}{i} lines, and the corresponding \Vsini\ values obtained from the interpolation in the grid. The last column shows the average \Vsini\ value and the corresponding dispersion computed from the \Vsini\ obtained from the individual lines.


\section{Discussion}\label{Discussion}

\subsection{Comparisons with the literature}\label{lit}

Our initial sample comprises 330 stars from eight open clusters in Carina. 
Some early-type stars of the studied open clusters
have published results of previous spectroscopic analysis.
Among these, 108 stars ($\sim$33\%) have at least one \Vsini\ measurement available in the literature, while 222 stars ($\sim$67\%) lack such data. Consequently, this study provides the first \Vsini\ measurements for approximately two-thirds of the sample.

\citet{Huang2006a, Huang2006} obtained the stellar parameters \teff, \logg, and \Vsini\ for a large sample of early-type stars in 19 open clusters, including Trumpler  14 and Trumpler  16, based on a method that combines fits of synthetic profiles of Balmer lines and \ion{He}{i} lines. \citet{Hanes2018} obtained \teff, \logg, and \Vsini\ for a sample of hot stars in Carina using non-LTE synthesis of the  \ion{He}{i} lines at  $\lambda\lambda$4026, 4388, and 4471 \AA. 
 
 NGC 3293 is a well-studied cluster with a larger number of stars having previous spectroscopic analysis.  \citet{McSwain} obtained  stellar parameters  \teff, \logg, and \Vsini\ from non-LTE synthesis based on the grid of Non-LTE models BSTAR2006, combining fits of synthetic profiles of $H\gamma$ and \ion{He}{i} lines at  $\lambda\lambda$4388, 4471 \AA\ and \ion{Mg}{ii} $\lambda$4481 \AA.
The method of \citet{Hunter} to obtain the parameters \teff\ and \logg\ for stars in NGC 3293  uses silicon ionization balance (\ion{Si}{ii} and \ion{Si}{iii}) and hydrogen lines, while  \Vsini\ is estimated based on  \ion{He}{i}  and metal lines. \citet{Morel} studied  160 stars with spectral types B1 to B9.5 in the cluster NGC 3293 in the context of the Gaia-ESO Survey, based on GIRAFFE and UVES spectra. They obtained  \teff, \logg\, and \Vsini\ using a global least-square minimization to derive the stellar parameters.

Other working groups of the Gaia-ESO Collaboration also determined \Vsini\ using different methods, such as 
 a combined Fourier transform plus a goodness-of-fit
methodology (The IAC Node) and  Non-LTE spectral synthesis (The Liège Node). Figure 8 of ~\citet{Blomme} shows the comparison between the \Vsini\ values obtained by different nodes of the Gaia-ESO Collaboration, including our node (ON Node). There, such a comparison was done for a sample of stars in common in the different studies including benchmark stars and other stars not belonging to the Carina region.  The comparison between the \Vsini\ values derived by the ON Node and those from the IAC and Liège Nodes shows good agreement.   \citet{2025arXiv250116508B} obtained rotational velocities for a sample of 38 O9-type and 27 B0-type stars in Carina, using {\tt iacob-broad} \citep{iacob_broad}, based on a method that combines Fourier Transform and the Goodness-of-fit to determine  the \Vsini\ from diagnostic lines, such as  \ion{He}{i} or metal lines.

\begin{figure*}[]
  \includegraphics[width=0.48\textwidth]{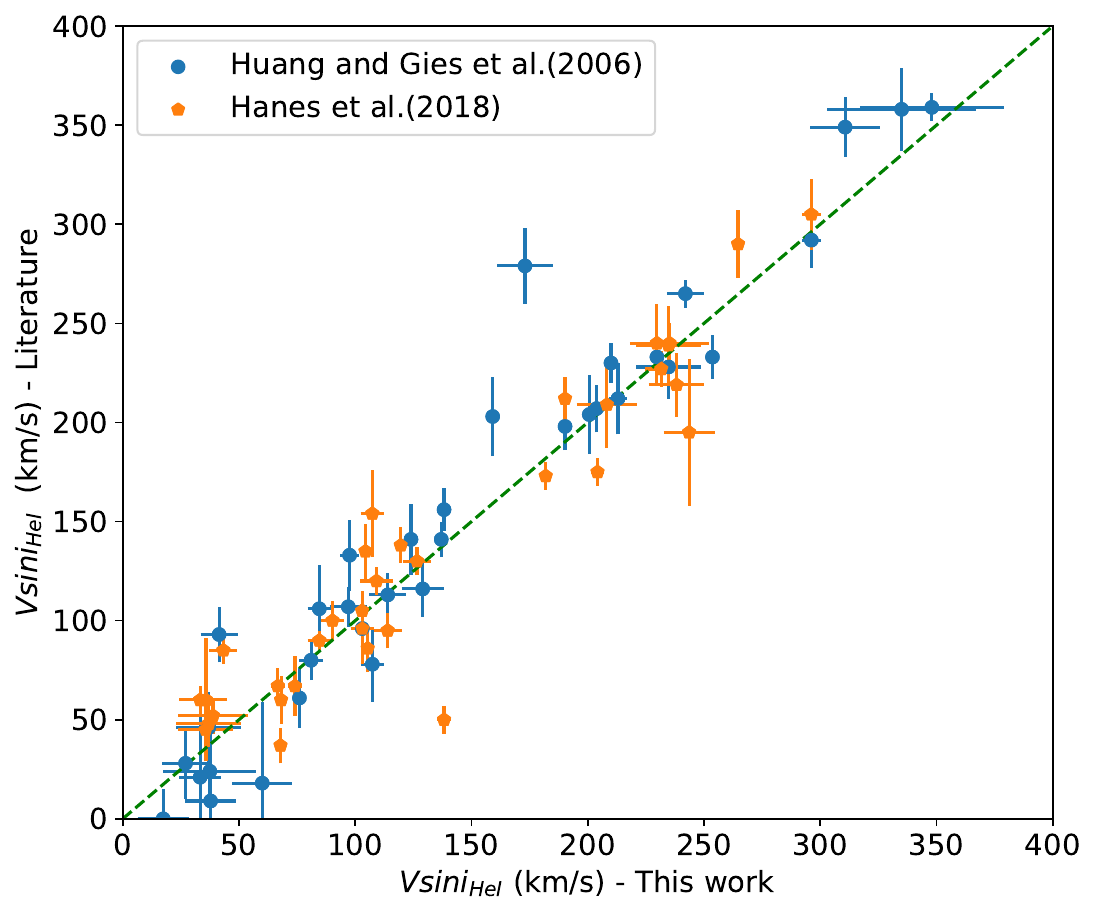}
  \includegraphics[width=0.48\textwidth]{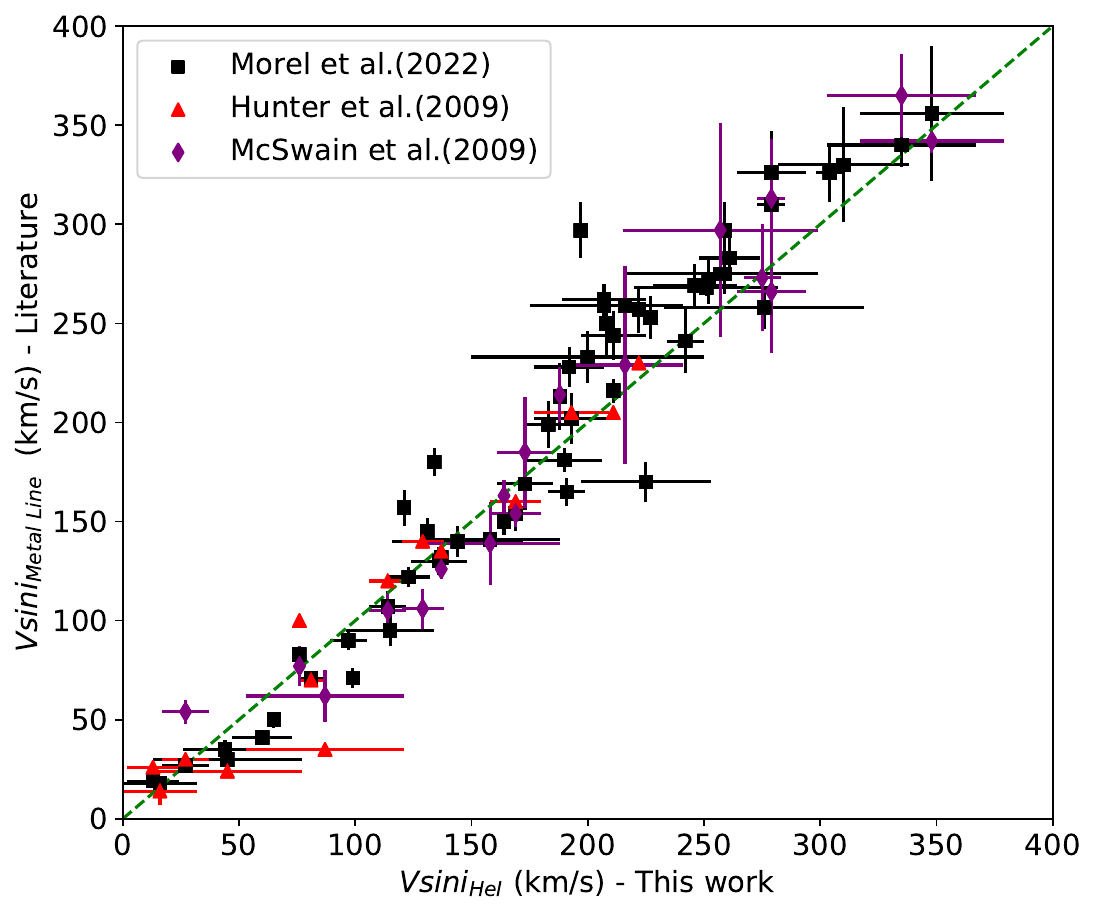}
    \includegraphics[width=0.48\textwidth]{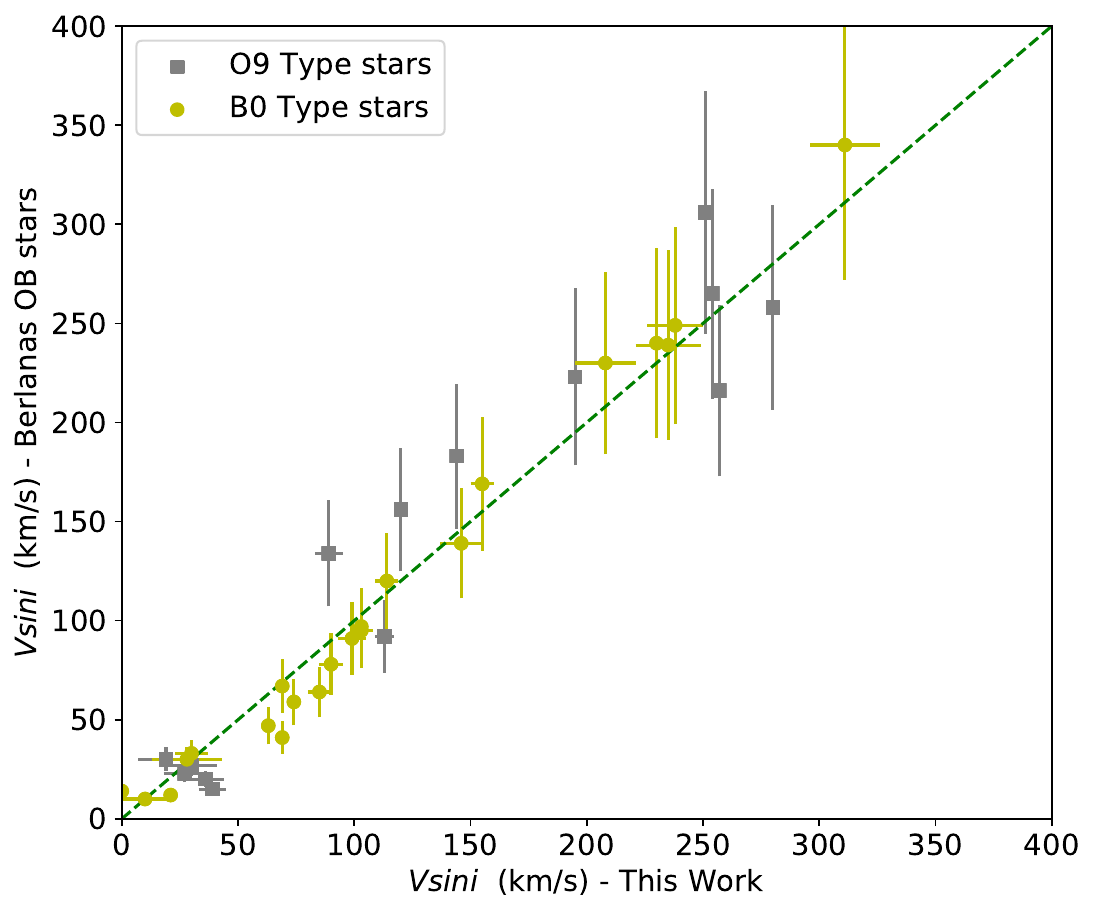}
  \includegraphics[width=0.48\textwidth]{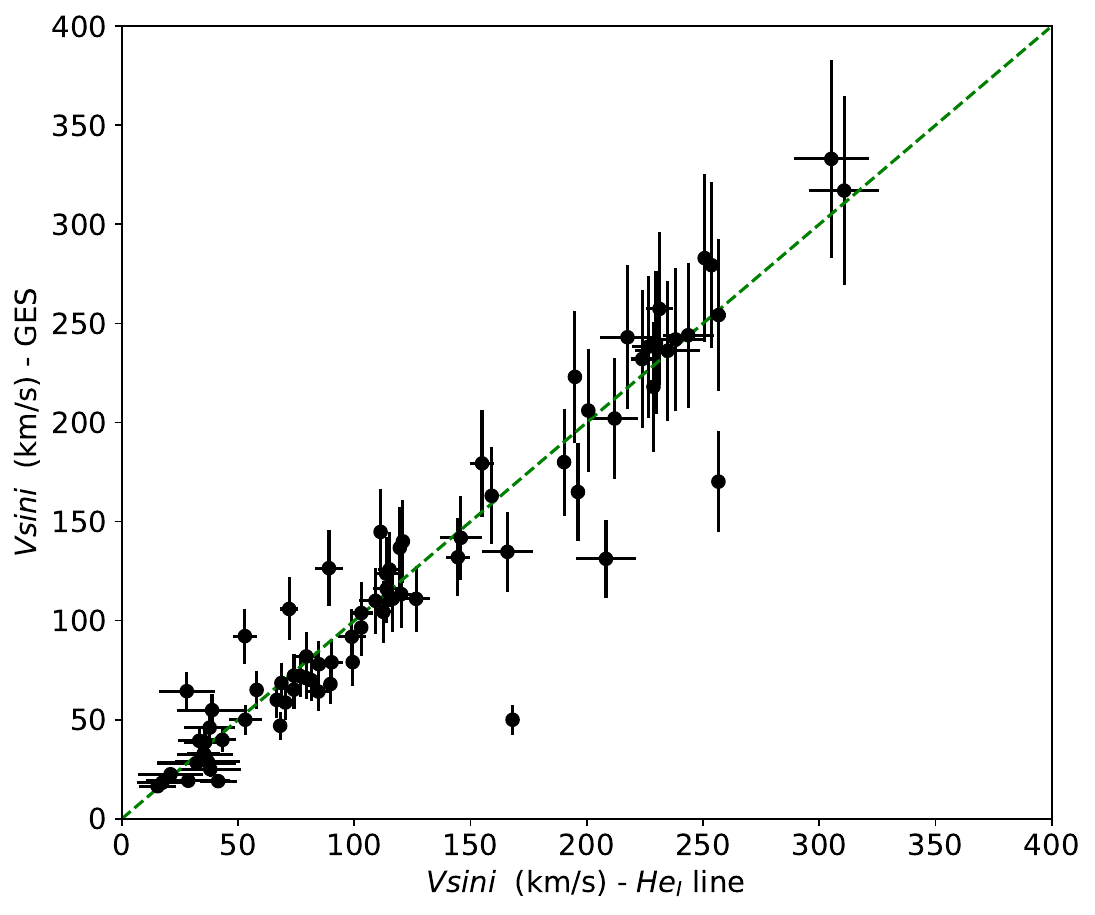}
  \caption{
Comparison of \Vsini\ values obtained in this study with literature results. The top-left panel presents published \Vsini\ values derived from the widths of \ion{He}{I} lines by \citet[][blue circles]{Huang2006} and \citet[][orange pentagons]{Hanes2018}, showing an average difference of $-3.9 \pm 25.2$ \kms\ relative to our measurements. The top-right panel displays \Vsini\ values obtained from spectral fits of helium and metal lines by \citet[][green triangles]{Hunter} and \citet[][purple diamonds]{McSwain}, with average differences of 1.9 and 3.2 \kms, respectively. Additionally, \citet[][black squares]{Morel} reported \Vsini\ values that are systematically lower than ours by 4\% for \Vsini\ $<$ 200 \kms\ and approximately 12\% higher for \Vsini\ $>$ 200 \kms.
The bottom-left panel compares our results with those of \citet{2025arXiv250116508B}, where \Vsini\ values for O9 stars (gray squares) exhibit slightly higher dispersion, while those for B0 stars (yellow circles) remain consistent with our estimates within the uncertainties. Finally, the bottom-right panel shows \Vsini\ values obtained by other Nodes of the Gaia-ESO Collaboration, with an average difference of $1.8 \pm 24.8$ \kms\ relative to our results. In all panels, the dashed line represents the one-to-one correlation (x = y).
  }
  \label{comp.lit}
  \end{figure*}

In Figure ~\ref{comp.lit} we compare our \Vsini\ results for each star with those found in the literature and results from other nodes of the GES Collaboration. The top left panel shows the comparison with the results based on the widths of \ion{He}{i} lines.  The points are color-coded according to the reference: results from \citet{Huang2006} are represented as blue circles and orange pentagons are from \citet{Hanes2018}.
The average difference in \Vsini\ between our results and the results of \citet{Huang2006,Hanes2018} obtained from the \ion{He}{i} lines is $-3.9\pm$25.2 \kms. 
The comparison between our results and those based on the fitting of metal and helium lines is shown in the top right panel. The averaged differences between our results and those of \citet[][red triangles]{Hunter} and \citet[][purple diamonds]{McSwain} are 1.9 and   3.2 \kms, respectively, in excellent general agreement within the uncertainties.
The results obtained by \citet[][black squares]{Morel} present different behavior according to the regime of stellar rotation:  for \Vsini\ $<$ 200 \kms, results by \citet{Morel}  are systematically lower than ours by a factor of 4\%, while for \Vsini\ $>$ 200 \kms, Morel's results are $\sim$ 12\%\ larger than ours. 
This behavior may be due to the relative importance of other broadening effects considered in the computation of the synthetic metal lines, as suggested by \citet{Daflon2007}.

The bottom-left panel compares our \Vsini\ estimates with those obtained by \citet{2025arXiv250116508B} for 14 O9 stars (gray squares) and 22 B0 stars (yellow circles) that overlap with our sample. The \Vsini\ results for the B0-type stars are consistent with our estimates within the uncertainties, while the results for the O9 stars show slightly higher dispersion.
Finally,   the bottom right panel of Figure~\ref{comp.lit} shows the comparison between our \Vsini\ estimates and those values obtained by other GES Nodes, considering only the stars in the Carina Nebula.  
The average difference between the two datasets is  
1.8$\pm$24.8 \kms.
The \Vsini\ values of most of the stars agree well within the uncertainties, with a few discrepant values,  
and the error bars tend to be larger for the higher \Vsini, which is expected and consistent with uncertainties expressed in relative difference.

\subsection{Stellar rotation and cluster age}

 \citet{Huang2006a} examined a sample of 496 early-type stars in 19 open clusters with varying ages, ranging from   6 to 72 Myr, with a median age of 12.5 Myr. They obtained an average value of $<$\Vsini$>=147\pm$26~\kms, 
 which matches  the average \Vsini\  for our sample,  $<$\Vsini$>$=159$\pm$91~\kms. In contrast to the sample of   \citet{Huang2006a},  the clusters in our study are younger, with a narrow range of $\log(age)$ between 6.678 and  7.139, as listed in table \ref{ages}. The adopted ages for the clusters Trumpler 14, Trumpler 15, Trumpler~16, Bochum 11 and NGC 3293 are from  \citet{Dias2021}, obtained by fitting theoretical isochrones to Gaia DR2  photometric data, as described in \citet{monteiro}. The ages of the clusters  Collinder 228 and Collinder 232 are from \citet{Loktin}, based on 2MASS photometric data.

 \begin{table}
  \centering
  \caption[]{\label{ages}
  {The Clusters ages}}
  \begin{tabular}{lcc} \hline\hline
Cluster & $\log(age)$  &  Ref. \\  
\hline
Trumpler 16~E/W & 6.678 $\pm$ 0.053 & \citet{Dias2021} \\
Bochum 11 & 6.721 $\pm$ 0.101 & \citet{Dias2021} \\
Trumpler 14  & 6.726 $\pm$ 0.006 & \citet{Dias2021} \\
Trumpler 15 & 6.814 $\pm$ 0.068 & \citet{Dias2021}  \\
Collinder 228  & 6.865 $\pm$ 0.032 & \citet{Loktin}\\
Collinder 232  & 6.874  &  \citet{Loktin}\\
NGC 3293  & 7.139 $\pm$ 0.016 & \citet{Dias2021}\\
\hline
  \end{tabular}
  \end{table} 

\begin{figure}
   \centering
   \includegraphics[width=\hsize]{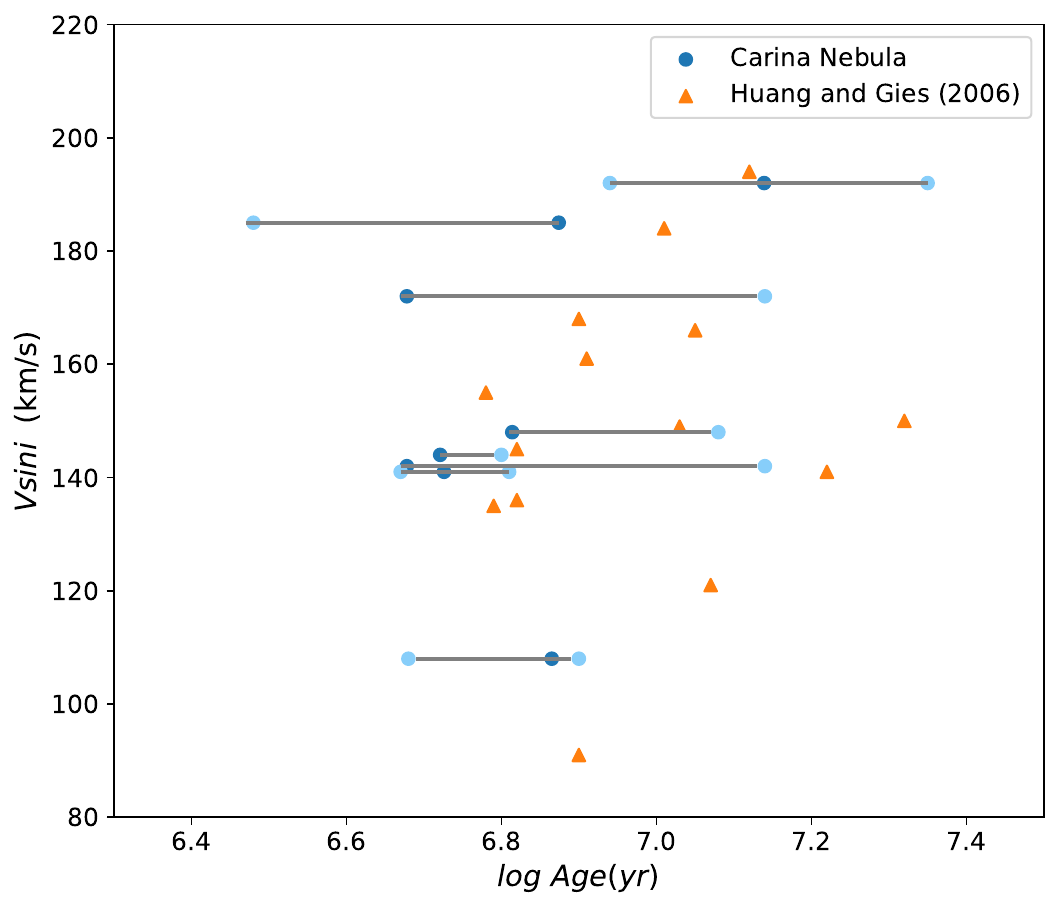}
  \caption{The median \Vsini\ of our open clusters as a function of the cluster age. The blue circles represent the data for the Carina clusters studied in this paper (dark blue for ages listed in table \ref{ages} and light blue for different ages available in the literature) and the orange triangles represent  the 14 open clusters studied by \citet{Huang2006a}.}
\label{age}
\end{figure}

 In Figure~\ref{age}, we plot the average \Vsini\ of young clusters as a function of the cluster age. Results of  \citet{Huang2006a} for fourteen  well-sampled clusters are represented by orange triangles. Our median \Vsini\ values for the Carina clusters are represented by the blue circles: dark blue circles represent the adopted ages listed in table \ref{ages} while light blue circles represent the different ages found in the literature for the studied clusters, connected by the horizontal lines. The youngest Carina cluster of our sample is Trumpler 16, which generally is not split into E and W components in the literature. The age of Trumpler 16, as one cluster,  
 varies from $\log(age)$ = 6.90 \citep{Conrad_2017} to 7.13 \citep{Cantat-Gaudin_2020}, which are all greater than the age we adopted, from \citet{dias21}. 
 Two clusters of our sample have scarce age estimates in the literature: the age of Bochum 11 obtained by  \citet{Cantat-Gaudin_2020},  $\log(age)$ = 6.80, is quite consistent with the age we adopted in our analysis, while the age of  Collinder 232 obtained by \citet{Preibisch} is $\sim$ 0.4 dex smaller than the value we list in table ~\ref{ages}. The logarithm of the ages of Trumpler 14 are consistent in many studies, for example,  6.67   \citep{Conrad_2017}.   and 6.81 \citep{huntreffert}. On the other hand, the  $\log(age)$ of Trumpler 15 obtained by \citet{Dias2021} is smaller than other values from the literature, such as 6.95 \citep{Cantat-Gaudin_2020} and 7.08  \citep{Conrad_2017,huntreffert}. Conversely, the age of Collinder 228 listed by \citet{Loktin} is greater than 
 6.68 by \citet{Conrad_2017} and 6.85 by \citet{huntreffert}. 
 The oldest Carina cluster of our sample is NGC 3293, with ages varying from  $\log(age)$ = 6.94 \citep{Conrad_2017} to 7.35 \citep{huntreffert}. The analysis of \citet{Morel} reported an age of 20 Myr for NGC 3293, which corresponds to $\log(age)$ = 7.30. 
 
  The discrepancy between the various age estimates of the studied clusters hampers the analysis of any possible trend of averaged \Vsini\  with age. Furthermore, our sample is unsuitable for this study, since all studied clusters are of very similar age, considering the uncertainty. Nevertheless, the largest number of fast-rotating stars is found in the oldest cluster of our sample, NGC 3293.

\subsection{\Vsini\ distributions}

\begin{figure}
   \centering
   \includegraphics[width=\hsize]{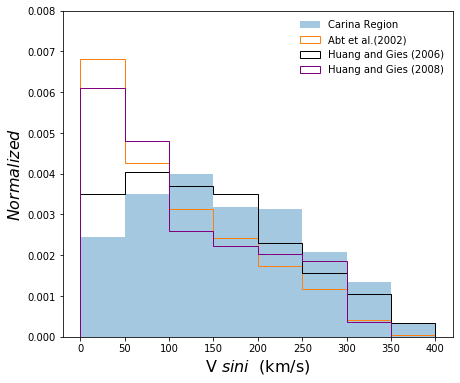}
      \caption{Distribution of  \Vsini\ obtained for  330 probable members of the Carina region (light blue histogram), compared to the distributions of \Vsini\ of early-type stars in clusters (dark blue line, \citet{Huang2006a}), and for field stars (\citet{Huang2008}; purple line) and in the field (orange line, \citep{Abt}).}
         \label{Dist.vsini}
   \end{figure}

The distribution of \Vsini\ for the sample of 330 probable members of the Carina region is shown as the light blue histogram in Fig. ~\ref{Dist.vsini},  with an average value of  $<$\Vsini$>$=159$\pm$91~\kms. 
The distribution of \Vsini \ has a peak between 100 - 150 \kms, which is a typical value for early-type stars (\citet{Wolff2007}; \citet{Wolff1982}; \citet{Huang2006a},  and \citet{Huang2006}). The distribution is consistent with the histogram of \Vsini\ for early-type stars in clusters (dark blue line, \cite{Huang2006}). 
Approximately 70 \% of the stars in our sample present \Vsini\ values higher than 100 \kms. 
On the other hand, the distributions of \Vsini\ for the field stars, represented by the orange and purple lines, \cite{Abt,Huang2008} respectively, have an excess of slowly rotating stars when compared to the stars in clusters.

\begin{figure}
   \centering
   \includegraphics[width=\hsize]{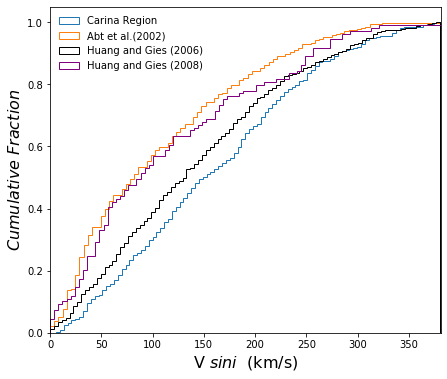}
      \caption{The cumulative fraction of stars as a function of \Vsini\  for different datasets of stars in clusters (blue: Carina, this work; black:  \citet{Huang2006a}) and in the field  (orange: \citet{Abt} and purple: \citet{Huang2008})}
         \label{cumulative.vsini}
   \end{figure}

In Fig. \ref{cumulative.vsini} 
we show the cumulative distribution functions (CDF) of 
\Vsini\ for our sample (blue line), as well as the CDFs for stars in clusters (black:  \cite{Huang2006a}) and for field stars   (orange: \cite{Abt}  and purple: \cite{Huang2008}).
The differences in the steepness of the curves provide insights into the rotational velocity distributions across different studies and regions. A steeper curve indicates a dataset with more stars having similar, lower \Vsini\ values. A more gradual curve indicates a broader range of rotational velocities, suggesting a more diverse population in terms of \Vsini.
The two CDFs for field stars are similar and both increase more steeply, particularly in the range of 0 to 200 \kms.
This suggests a higher concentration of stars with lower \Vsini\ values compared to the Carina Region. The curves for field stars reach the cumulative fraction of 1 more quickly, indicating fewer high-velocity stars.
A K-S test between our sample in Carina and the samples of field stars rejects the null hypothesis, assuming a confidence level of 95$\%$.
The CDF for stars in Carina increases more gradually compared to the other datasets, indicating a wider spread of \Vsini\ values.
The CDF for stars in clusters by   \citet{Huang2006a} displays a similar trend as the one for stars in Carina, although the former distribution is steeper than the Carina curve.  

These results align with the conclusions presented by  \citet{Wolff2007} and \citet{Huang2006} that early-type stars in clusters tend to show higher \Vsini\ values when compared to field stars.
Analyses by \citet{Abt, Huang2006} suggested that the lower concentration of slow-rotating OB stars in different density scenarios could be related to evolutionary processes. 
 
Field stars are generally older than stars in young clusters, as they may represent remnants of dissolved clusters \citep{giesreview,braganca2012}. Over time, processes such as magnetic braking \citep{stepien2000}, stellar wind-driven mass loss \citep{mem2000}, and tidal interaction in close binaries \citep{zahn1977} can reduce stellar rotation rates. These mechanisms may contribute to the lower \Vsini\ values observed in field populations. This supports the interpretation of field stars having, on average, slower rotation due to their more advanced evolutionary stage and cumulative angular momentum loss.

However, \citet{Wolff2007} suggests that this relationship is not the most appropriate, proposing that star clusters that survive the bound cluster formation process likely originate from dense environments. These OB stars typically exhibit characteristics such as a high average accretion rate during stellar formation, short core collapse times, and early disk disruption through photoevaporation. These factors may contribute to the emergence of rapid rotators within denser clusters.
Conversely, objects that form in unbound associations are thought to originate from regions with lower surface density and shorter accretion times. Lower surface density can result in weaker interactions with neighboring objects, making these stars less susceptible to early disk disruptions. This could help preserve a higher stellar rotation rate.

  \citet{Wolff2007}  investigated the influence of environmental density on the \Vsini\ distribution in clusters. This study included eight clusters formed in high-density regions, seven low-density clusters, and field stars, suggesting that denser environments tend to produce more fast-rotating stars.  
  \cite{Braganca2019} studied the rotation of  350 B-type stars located in different environments of the  Galactic disk, including field stars and OB associations. Their results suggest that the distribution of \Vsini\  of the field stars exhibits a significant concentration at lower values (\Vsini\ < 50 \kms).
Conversely, the peak of the \Vsini\ distribution for stars in OB associations is shifted towards higher values, between  50 and 100 \kms. In the context of \citet{Braganca2019}'s study, our results suggest that there is a general trend of finding more rapidly rotating stars in the open clusters in Carina when compared to field OB stars \citep{Abt, Huang2008}, which is consistent with the trend previously identified by \citet{Wolff2007} and \citet{Braganca2019}.

  To better understand the relationship between the density of the clusters in Carina and their rotation rates, it would be necessary to estimate the density of the clusters. However, significant reddening in the Carina region, which deeply affects the completeness in the region (Section \ref{member}), prevents a study of \Vsini\ in terms of the densities of the Carina clusters.

\begin{figure}
   \centering
   \includegraphics[width=\hsize]{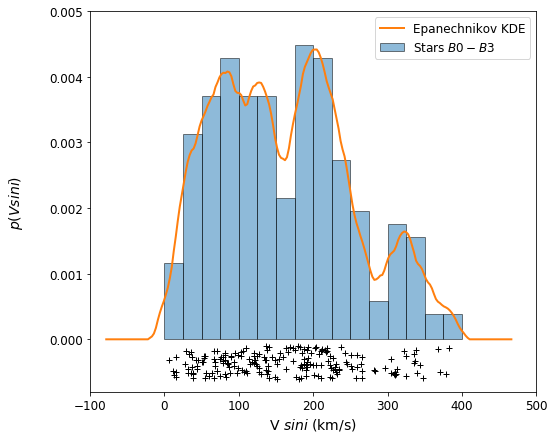}
      \caption{Distribution of \Vsini\ for 206 B0-B3 stars of our sample. 
      The PDF obtained with a Epanechnikov kernel is represented by the orange line and the black symbols at the bottom represent the \Vsini\ estimates in each bin.   
       The vertical scatter of points is artificial to avoid overlap and does not carry any quantitative meaning.
      The distribution is bimodal, with a third, smaller peak at high \Vsini.  }
         \label{hist_kde}
   \end{figure}

 \citet{Dufton13}  studied a sample of O9.5-B3 stars in the 30Dor region, in the context of the VLT-FLAMES Tarantula Survey for a sample of  334 young OB stars. 
 They used the Fourier Transform method to derive \Vsini\ and their obtained  distribution is bimodal, 
 with $\sim$25\% of the sample presenting \Vsini < 80 \kms. 
  Following \citet{Dufton13}'s approach, we segregate those stars in Carina from our sample with spectral types between B0 and B3. The adopted spectral types of the sample stars come from different sources listed in the SIMBAD Database. When spectral subtype was not available, we used the effective temperature and the relative intensity of key metal lines as a rough estimate of the spectral subtype. 
  Figure \ref{hist_kde} presents a histogram of \Vsini\ for this subsample of  206   B0-B3 stars,
  showing the frequency distribution of the \Vsini\ estimates,  represented by the black crosses at the bottom.     The vertical scatter of points is artificial to avoid overlap and does not carry any quantitative meaning.  Overlaid on the  histogram, the orange line represents  
  the overall probability density function (PDF) for \Vsini, fitted using an Epanechnikov kernel. The bandwidth was chosen according to the {\it rule of thumb} \citep{Silverman86}. The distribution shows two prominent peaks, suggesting a bimodal distribution of  \Vsini\ among the B0 to B3 stars. The first peak is centered around 100 \kms and the second peak is centered around 200 \kms. There are few stars with \Vsini\ values above 350 \kms, indicating these extreme velocities are rare in this sample. A similar statistical analysis was performed using a combination of Gaussian distributions, confirming the bimodality of the \Vsini\ distribution.
  
  In \citet{Dufton13},  the bimodal distribution has been interpreted as an indication of two different populations of stars within these spectral types, resulting from different formation conditions,  stellar ages, or even the presence of nearby companions. However, \citet{Dufton13} found no evidence of a correlation between the stars's rotation and their relative positions in the 30Dor region.   The OB stars of our sample have been formed in clusters within the large complex of Carina association, with similar ages and distances \citep{Goppl2022}. The full comprehension of the big picture in Carina still lacks more information about the stellar populations, such as chemical composition, that could be obtained from a detailed spectroscopic analysis.

  \citet{Morel} studied approximately 160 stars in NGC 3293, with spectral types between B1 and B9.5. They derived stellar parameters, \Vsini,  and chemical abundances based on non-LTE synthesis. The \Vsini\  measurements were convolved with synthetic spectra for all GIRAFFE configurations and the best fits were chosen using $\chi^2$.  The distribution of the apparent \Vsini\  for their sample is Gaussian, with a peak at \Vsini\ around 250 \kms. 
  \citet{Morel} found no evidence of a bimodal structure in the \Vsini\ distribution of  NGC 3293 and they suggest this may be because their sample includes stars cooler than B3.  Our analysis of NGC 3293, based on a sample of 66 stars with earlier spectral types, between O9.5 and B6, also results in a Gaussian \Vsini\ distribution, although its peak is shifted towards slightly lower \Vsini\ $\sim$  200 \kms.

  At the age range of the oldest open cluster of our sample, NGC 3293, binary interaction may be responsible for the enhanced rotational velocity distribution.
 As shown by \citet{mcswain2005}, many Be stars are spun up through binary mass transfer mechanisms. The larger number of fast rotators in NGC 3293 may thus reflect evolutionary processes in close binaries that have had enough time to operate. This interpretation aligns with the observed peak in \Vsini\ distribution and supports the idea that Be stars emerge preferentially in slightly older stellar populations due to post-interaction spin-up effects. We note, however, that the census of Be stars in the sample clusters was not an objective of this work.

\citet{berlanas23} studied the population of OB stars in the same Carina clusters studied in this paper, but not NGC 3293.  Their sample comprises stars with spectral types earlier than B2 in the case of dwarfs, earlier than B5 in the case of giants, and all spectral types in the case of supergiants. The spectra of the B stars are mainly from Gaia-ESO Survey while the data for the O stars are from the GOSSS project \citep[][and references therein]{Maiz2016}. \citet{berlanas23} discussed the distribution of stellar rotation in terms of a rotation index that qualitatively indicates the line broadening, from slight to severe.  For this sample of O and early-B types stars, they obtained a distribution that peaks at the bin 200 - 250 \kms, suggesting 
their sample presents a higher concentration of rapidly rotating stars compared to the distributions of the clusters studied in this work.  This may result from the fact that our sample includes stars with spectral types between B3 and B6 that are not present in Berlanas' sample.

 \begin{figure}
   \centering
   \includegraphics[width=\hsize]{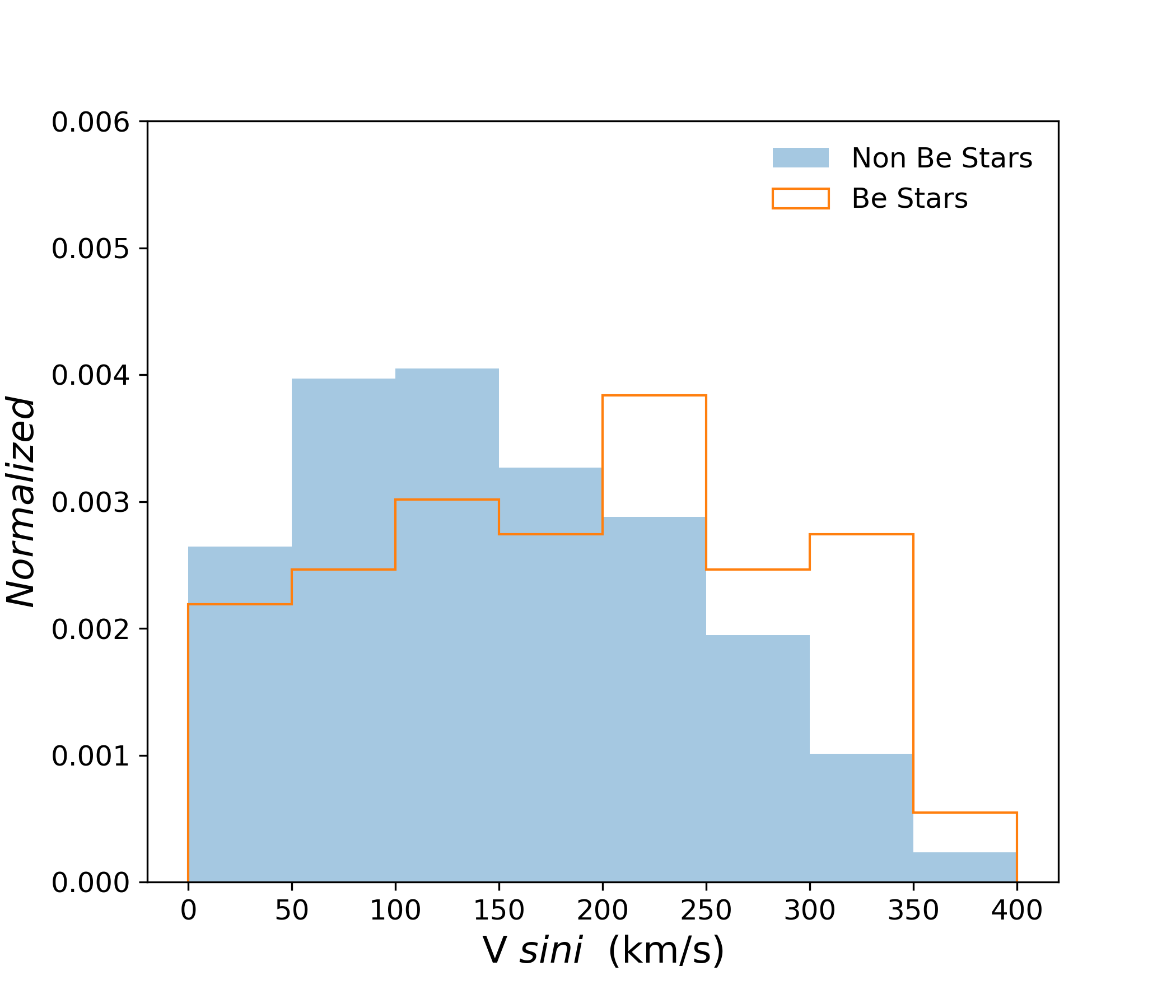}
      \caption{ Distributions of \Vsini\ for  73 Be (orange) and  257 Non-Be (blue) stars in Carina. 
      The peak of the  \Vsini\ distribution for the Be stars is shifted towards higher values. }
         \label{hist_Be}
   \end{figure}

 Stellar rotation is believed to play a crucial role in the formation of circumstellar disks observed around B-type stars with emission lines in their spectra—commonly known as Be stars \citep{RCM2013}. Some stars in our sample exhibit emission lines in their observed spectra. However, as mentioned in Section~\ref{member}, the Carina region is embedded in dense molecular clouds, so some central emissions may have a nebular origin. 

The observed emission lines are classified based on visual inspection of their profiles and measurements of the full width at half maximum (FWHM). Narrow, single-peaked lines are classified as nebular emission, and the corresponding sources are considered non-Be stars. In contrast, broad emission lines — sometimes displaying double peaks or P-Cygni profiles — are likely formed in circumstellar disks.
Based on this classification approach, 73 stars in our sample were identified as Be stars, and 257 as non-Be stars (including those with probable nebular emission lines). 
 
We note however the number of stars classified as Be stars may be overestimated, especially in Trumpler 14 and Trumpler 16W,  as some of the detected emission lines could originate from nebular rather than from a rotating circumstellar disk. A detailed analysis to confirm the origin of these emission profiles is beyond the scope of this paper.

 Figure~\ref{hist_Be} shows the \Vsini\ distributions for non-Be stars (blue histogram, average \Vsini = $151 \pm 88$ \kms) and Be stars (orange histogram, average \Vsini\ of 187$\pm$99 \kms). Be stars, in particular, tend to rotate more rapidly, with a higher concentration of fast rotators and a distribution peak shifted slightly toward higher \Vsini\ values. These results are consistent with previous studies indicating that Be stars are generally rapid rotators, often approaching their critical rotational velocities. It is important to note that \Vsini\ represents the projected rotational velocity along the line of sight; thus, the true equatorial velocities are likely higher, especially for stars observed at lower inclination angles.

\begin{figure}
   \centering
   \includegraphics[width=\hsize]{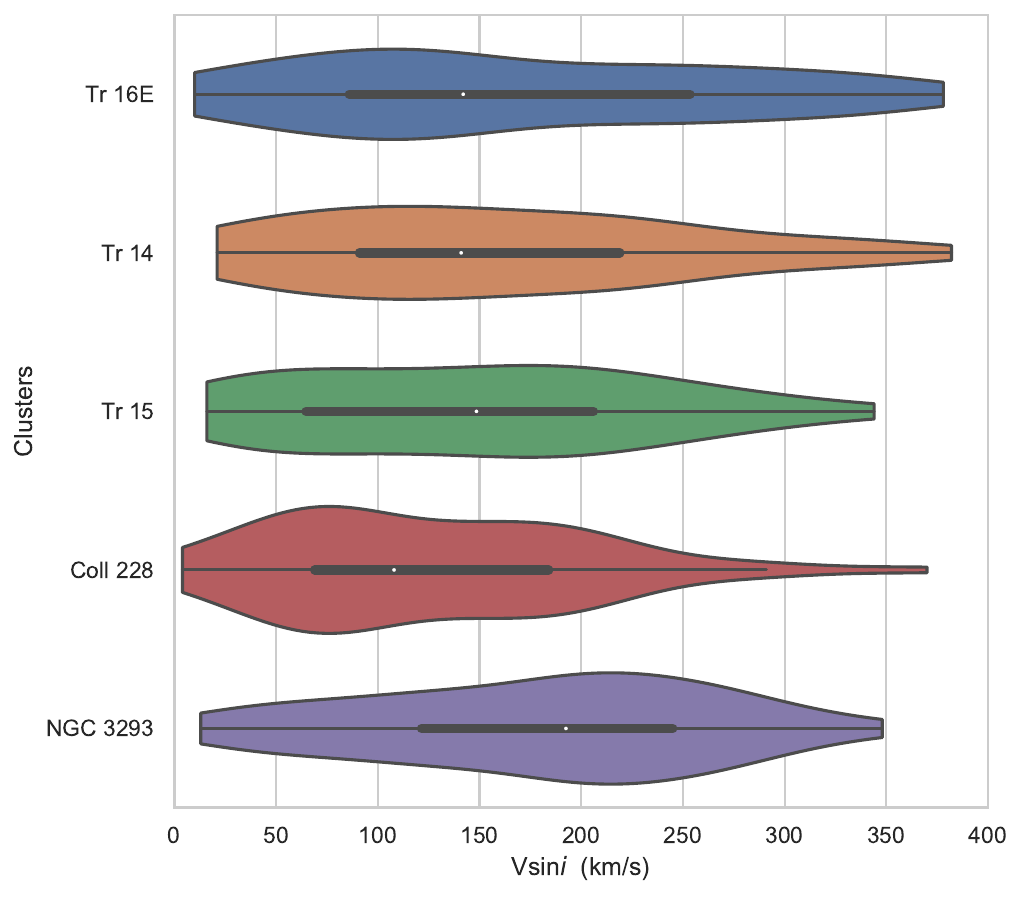}
      \caption{Violin distributions of  \Vsini\ of early-type stars of the open clusters Trumpler~16~E (blue) Trumpler 14(orange), Trumpler 15 (green), Collinder 228 (red) and NGC 3293 (purple).  The distributions are arranged according to increasing clusters ages, from top to bottom. White dots in the plots indicate the median, while the thick bars represent the interquartile range and the thin bars show the 95\% confidence interval. Broader regions of the distribution represent a higher probability that a star will have that \Vsini\ value.}
         \label{vsini_violin}
   \end{figure}

Figure \ref{vsini_violin} presents a violin plot showing the distribution of \Vsini\ 
values of open clusters with more than 40 stars: Trumpler 14, Trumpler 15, Trumpler~16~E, Collinder 228, and NGC 3293 \footnote{The distributions of \Vsini\ in the clusters  Trumpler~16~W, Bochum 11, and Collinder 232 are not discussed here due to the small number of stars in these three clusters.}.
This type of plot is useful for comparing data distributions across multiple groups simultaneously. A colored violin plot represents each cluster in the figure,  a combination of a box plot and a kernel density plot. They show the distribution of the data across different values: The width of the violin plot at different values of \Vsini\ indicates the density of stars with those velocities; the white dot in the center of the black bar represents the median value of \Vsini\ for that cluster. The black bar in the center of each violin represents the interquartile range (IQR), which contains the middle 50\% of the data, and the thin black line extending from the black bar represents the range of the data, excluding outliers.

The violin plot for Trumpler  14 shows a median \Vsini\  of 141$\pm$95 \kms,  and a fairly symmetric distribution, with moderate spread, indicating a balanced distribution of rotational velocities. The median \Vsini\ for Trumpler 15  is  148$\pm$88 \kms\ and the distribution is narrower compared to Trumpler 14, indicating less variations in rotational velocities within this cluster. The distribution of \Vsini\ for Trumpler  16E  is quite narrow with a lower spread, showing a high concentration of stars with similar rotational velocities and the median value is 142$\pm$104 \kms. Collinder  228 has a wider spread compared to previous clusters, indicating higher variability in rotational velocities, and the median \Vsini\ is  108$\pm$77 \kms.  Finally, NGC 3293 presents the highest median \Vsini\ among the clusters, 192$\pm$84 \kms, and the distribution is wide and asymmetric, with a long tail towards higher velocities, indicating a significant number of stars with high rotational velocities.

The violin plots 
for Trumpler 15 and Trumpler~16~E show less variation in \Vsini, indicating more homogeneous populations in terms of rotational velocity. In contrast, NGC 3293 has a broader and skewed distribution, suggesting a more diverse population with many stars rotating at higher speeds. 
The purple violin in Figure  \ref{vsini_violin} suggests that it might have a relevant contribution of NGC 3293 stars at the peak at \Vsini\ $\sim$ 200 \kms\ in Figure \ref{hist_kde}.
Collinder 228 also shows considerable variability in rotational velocities.  Our results suggest that, in general, stars in Collinder 228 tend to have lower \Vsini\ than the stars in NGC 3293. 
   
\section{Conclusions}\label{Conclusions}

We analyzed high-resolution spectra for a sample of O9.5-B6 stars members of 8 open clusters in the Carina Nebula: Trumpler 14, Trumpler 15, Trumpler~16~E, Trumpler~16~W, Collinder 228, Collinder 232, Bochum 11, and NGC 3293.   The observational data was provided by the Gaia$-$ESO Survey, a large public spectroscopic survey
that used UVES and GIRAFFE spectrographs.
The cluster members have been confirmed based on the distributions of astrometric and photometric data from Gaia, and radial velocities measured from helium and silicon lines. 

Projected rotational velocities (\Vsini) were derived from the widths of \ion{He}{i} lines for a sample of 330 early-type stars that are probable members of the Carina clusters. For 222 of these stars, \Vsini\ values are reported for the first time. Additionally, this study presents the first \Vsini\ estimates in the literature for members of the clusters Trumpler 15, Collinder 228, Collinder 232, and Bochum 11.
The distribution of  \Vsini\ for the full sample has a peak in the bin 100$-$150 \kms, which is consistent with typical distributions of \Vsini\ of stars in clusters, but differs from the distributions of field B stars, which generally show an excess of low \Vsini\ values. 
For a sub-sample of stars with earlier spectral types between B0 and B3, the \Vsini\ distribution turns into a bimodal distribution, with peaks at 100 and 200 \kms, plus a third, small peak at 350\kms. 

The \Vsini\ distributions of each cluster, considering only those clusters with more than 40 analyzed stars,  are slightly different in terms of the median \Vsini\ values and the distribution shape and symmetry. For example, the oldest cluster of our sample, NGC 3293, presents a wide, asymmetric distribution, with the highest median \Vsini\ among the clusters,  suggesting this cluster has a larger number of rapidly rotating stars. On the other hand, the distribution for Trumpler~16~E is quite narrow, indicating that the stars' \Vsini\ have a lower dispersion.  Collinder 228 presents the lowest median \Vsini\ among the Carina clusters,  126$\pm$77 \kms.


\section*{Acknowledgements}
We thank the referee for carefully reading our manuscript and for the constructive comments and suggestions.
My sincere thanks to my wife, F. Carine, for her constant partnership and support throughout the development of this research.
W.S acknowledges financial support from CAPES and FAPERJ  for Ph.D. fellowships.
      S.D. acknowledges CNPq/MCTI for grant 306859/2022-0 and FAPERJ for grant 210.688/2024. 
    T.M. acknowledges financial support from Belspo for contracts PRODEX Gaia-DPAC and PLATO mission development.
    S.R.B. and A.H. acknowledge funding from the Spanish Ministry of Science and Innovation (MICINN) through the Spanish State Research Agency through grants PID2021-122397NB-C21, and the Severo Ochoa Programme 2020-2024 (CEX2019-000920-S). S.R.B. also acknowledges financial support by  NextGeneration EU/PRTR and MIU (UNI/551/2021) through grant Margarita Salas-ULL.
    J.~M.~A. acknowledges support from the Spanish Government Ministerio de Ciencia e Innovación and Agencia Estatal de Investigaci\'on (10.13\,039/501\,100\,011\,033) through grant PID2022-136\,640~NB-C22 and from the Consejo Superior de Investigaciones Cientificas (CSIC) through grant 2022-AEP~005. 
    \\
    Based on observations collected at the ESO telescopes under programme 188.B3002, 193.B-0936, and 197.B-1074, the Gaia-ESO Public Spectroscopic Survey. These data products have been processed by the Cambridge Astronomy Survey Unit (CASU) at the Institute of Astronomy, University of Cambridge, and by the FLAMES/UVES reduction team at INAF/Osservatorio Astrofisico di Arcetri. These data have been obtained from the Gaia-ESO Survey Data Archive, prepared and hosted by the Wide Field Astronomy Unit, Institute for Astronomy, University of Edinburgh, which is funded by the UK Science and Technology Facilities Council.
    This work was partly supported by the European Union FP7 programme through ERC grant number 320360 and by the Leverhulme Trust through grant RPG-2012-541. We acknowledge the support from INAF and Ministero dell' Istruzione, dell' Universit\`a' e della Ricerca (MIUR) in the form of the grant "Premiale VLT 2012". The results presented here benefit from discussions held during the Gaia-ESO workshops and conferences supported by the ESF (European Science Foundation) through the GREAT Research Network Programme.


\section*{Data Availability}

The GES spectra analyzed in this paper are available at the GES Science Archive (http://ges.roe.ac.uk/). 
 



\bibliographystyle{mnras}
\bibliography{reference} 




\appendix

\section{Photometric and astrometric data}\label{cmds}

In this appendix, we present through Figures \ref{member_tr16e} and \ref{member_ngc3293} the distributions of astrometric and photometric data from Gaia EDR3, as well as radial velocities measured from the observed spectra, for the clusters Trumpler 15,  Trumpler~16~E, Trumpler~16~W, Bochum 11, Collinder 228, Collinder 232, and NGC 3293.

\begin{figure*}
\centering
\includegraphics[width=\hsize]{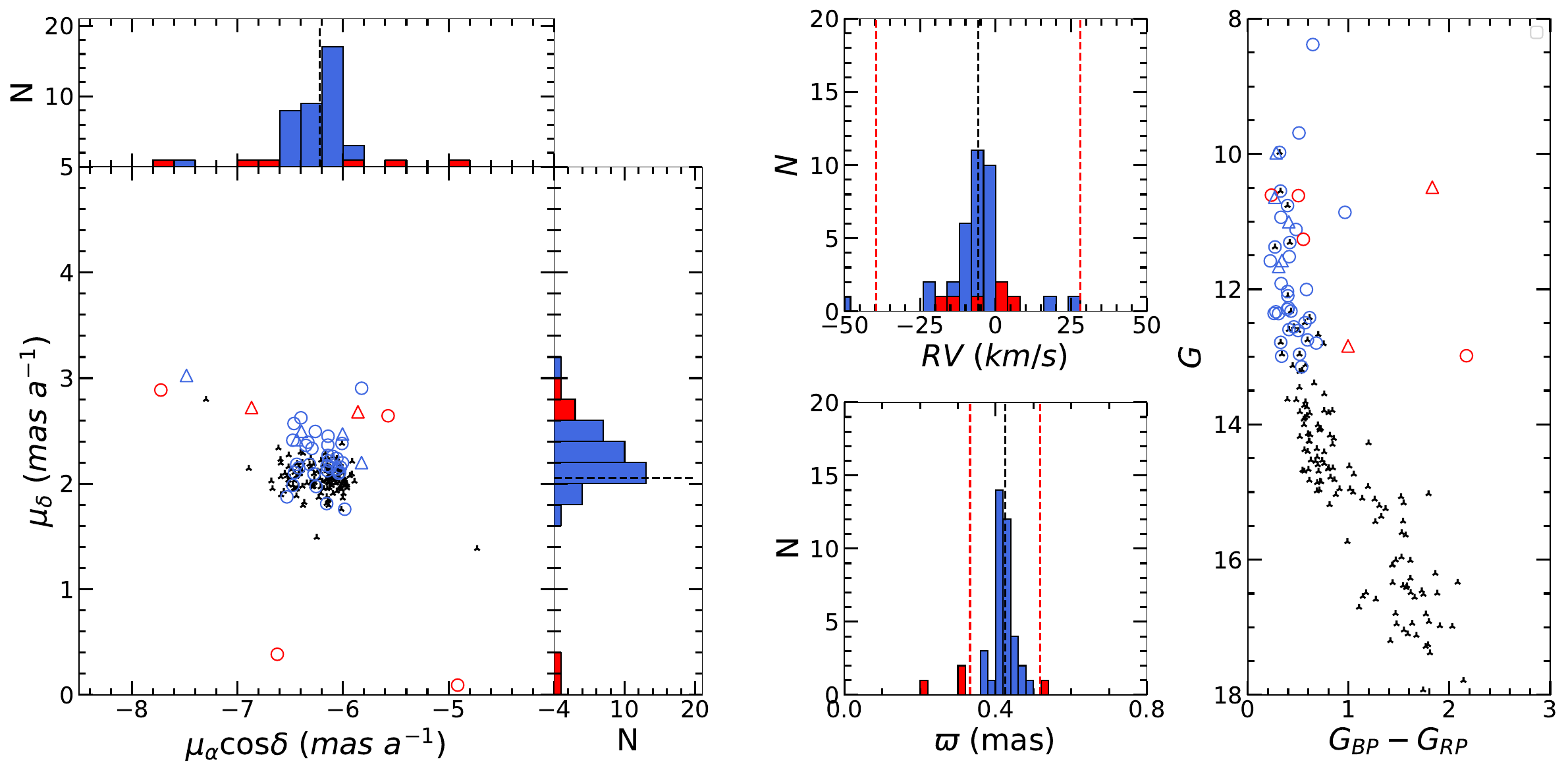}
\caption{Gaia EDR3 astrometric and photometric data for the cluster Trumpler~15. The small black triangles represent the cluster members by \citet{cantat}. Open symbols represent early-type stars with \Vsini\ measurement, with the blue circles representing the cluster members and the red circles representing non-members. The open triangles with the same color code refer to the stars selected for the abundance analysis, which will be presented in a forthcoming paper. 
{\it Left panel}: the distribution of proper motions in right ascension and declination. The side histograms are color-coded according to the symbols and the dashed line depicts the mean value for our sample. The middle panels show the distributions of the sample stars' radial velocities (upper) and parallax (lower),  color-coded according to the open circles. The black dashed lines are the mean values while the red dashed lines represent the 3$\sigma$ limits for the sample stars. The right panel presents the color $\times$ magnitude diagram using Gaia magnitudes.}
\label{member_tr15}
\end{figure*}

\begin{figure*}
\centering
\includegraphics[width=\hsize]{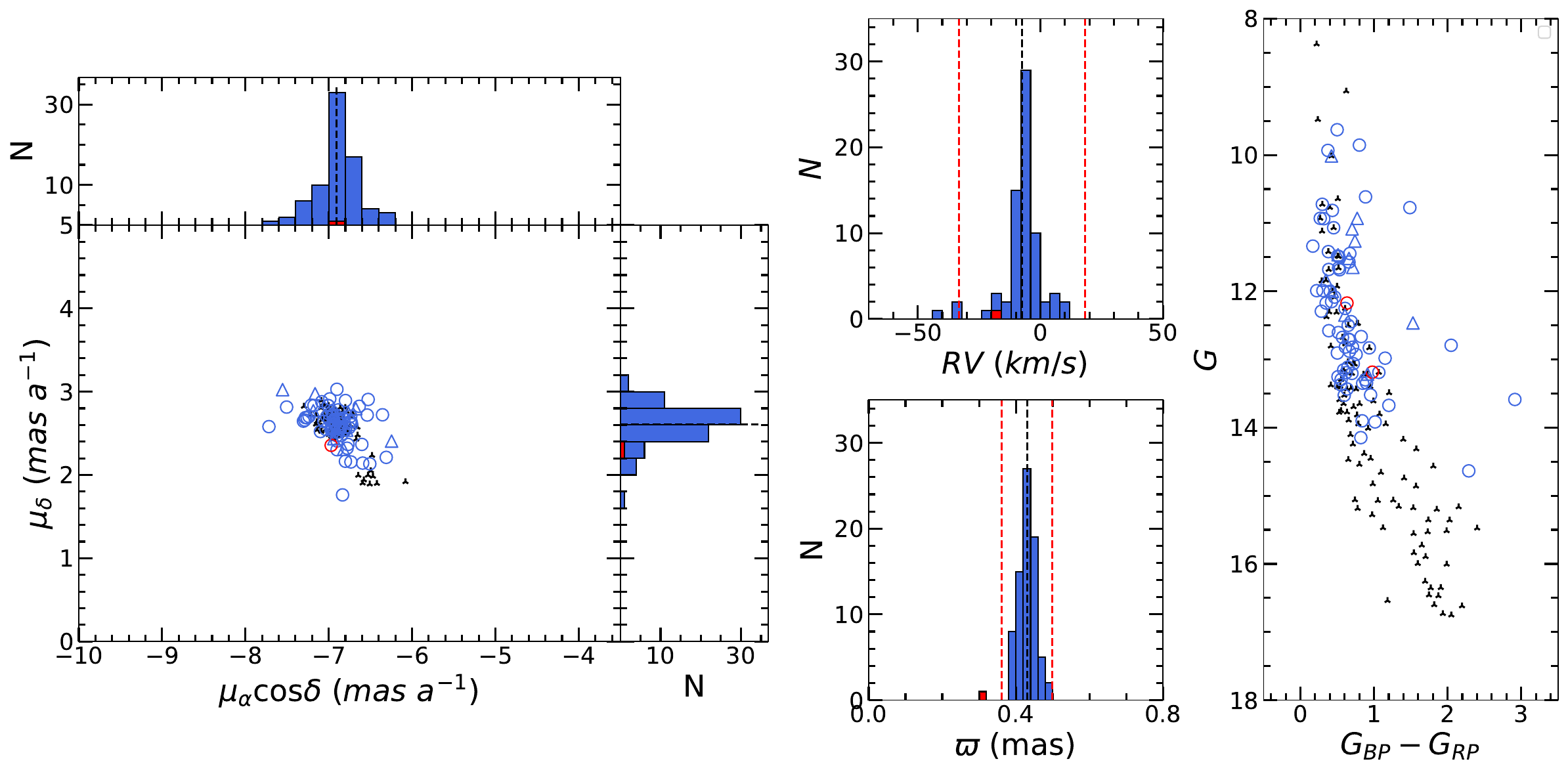}
\caption{Same as figure \ref{member_tr15} for the cluster Trumpler~16E}
\label{member_tr16e}
\end{figure*}

\begin{figure*}
\centering
\includegraphics[width=\hsize]{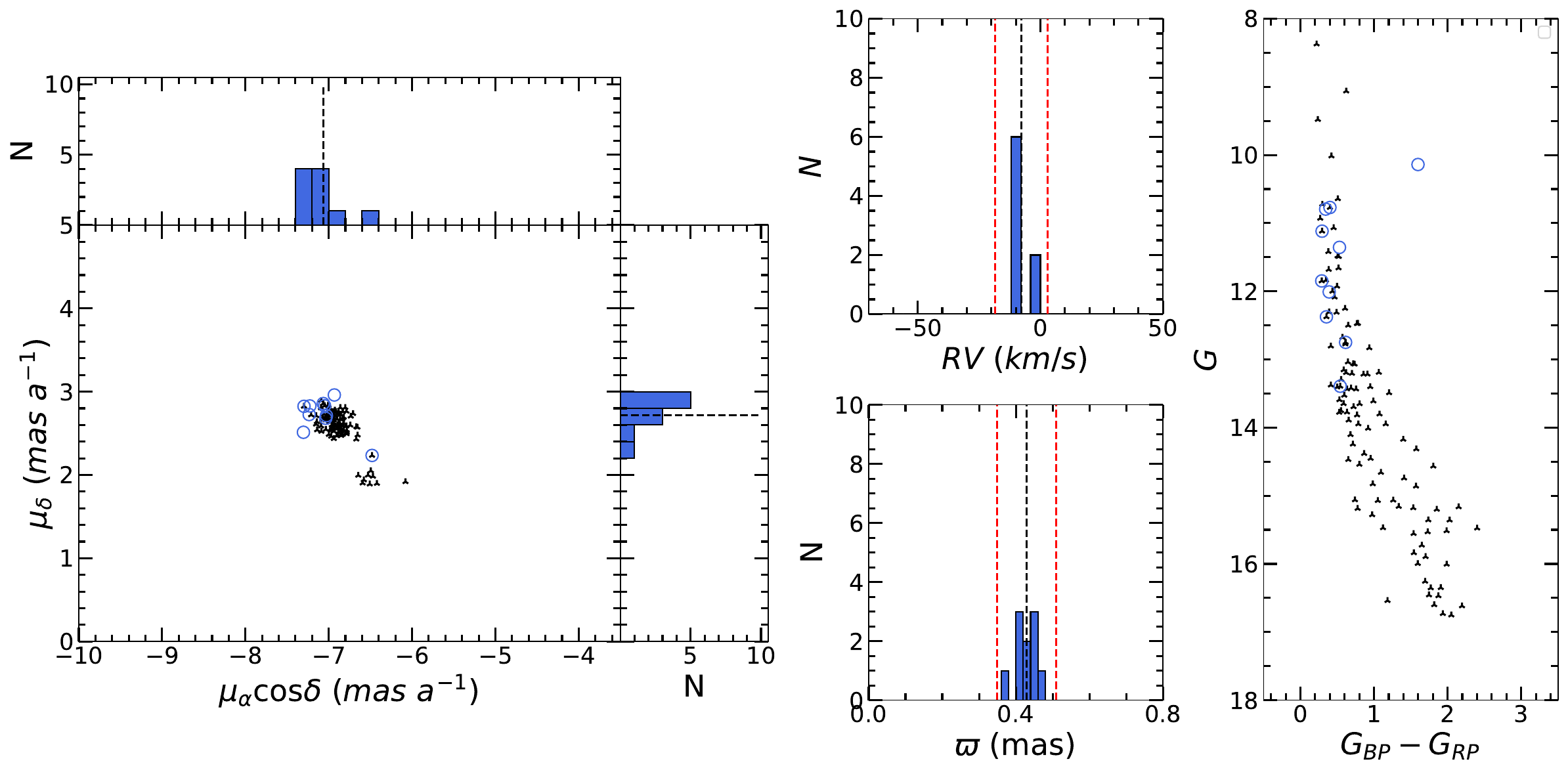}
\caption{Same as figure \ref{member_tr15} for the cluster Trumpler~16~W}
\label{member_tr16W}
\end{figure*}

\begin{figure*}
\centering
\includegraphics[width=\hsize]{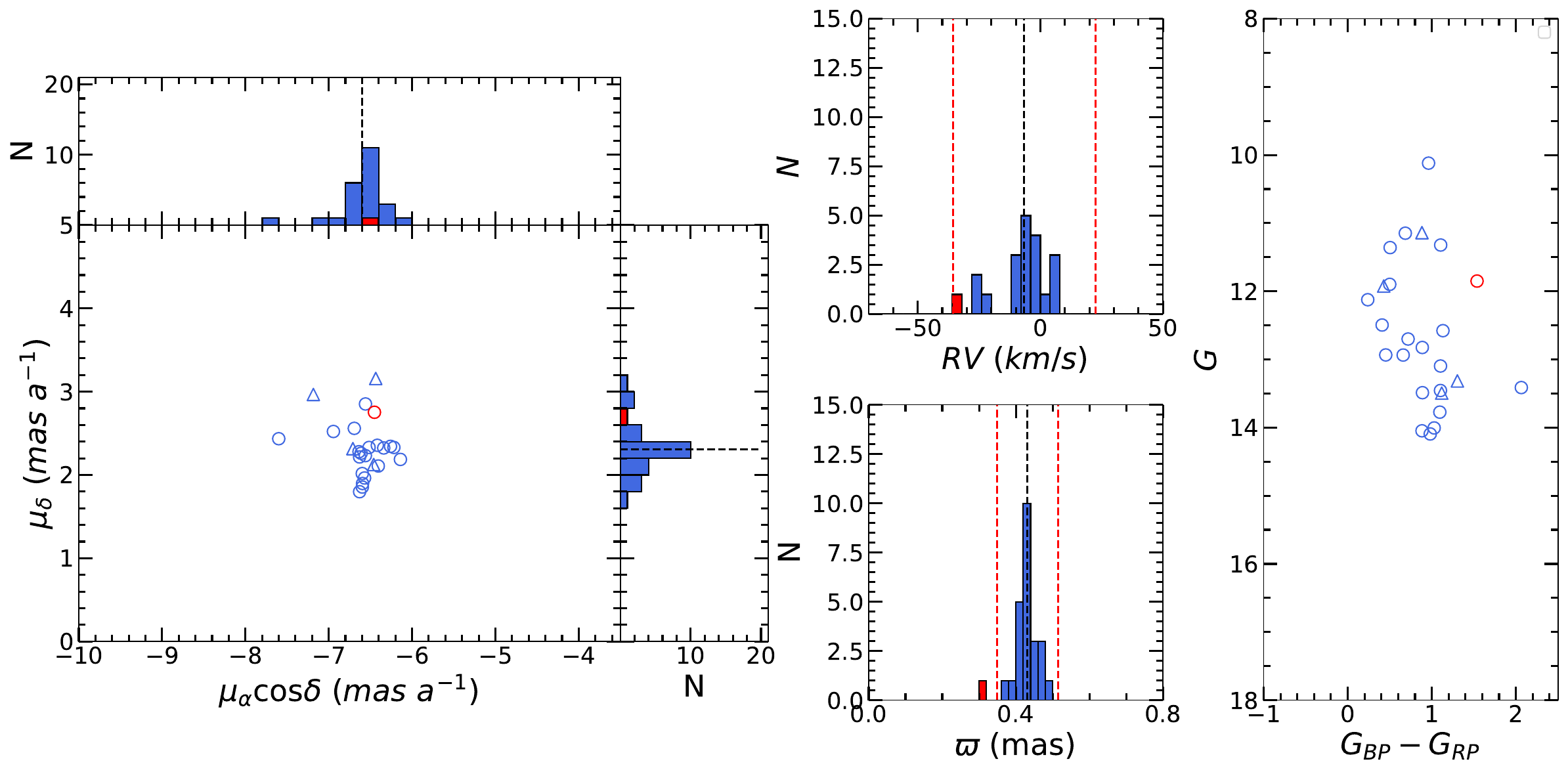}
\caption{Same as figure \ref{member_tr15} for the cluster Bochum 11}
\label{member_b11}
\end{figure*}

\begin{figure*}
\centering
\includegraphics[width=\hsize]{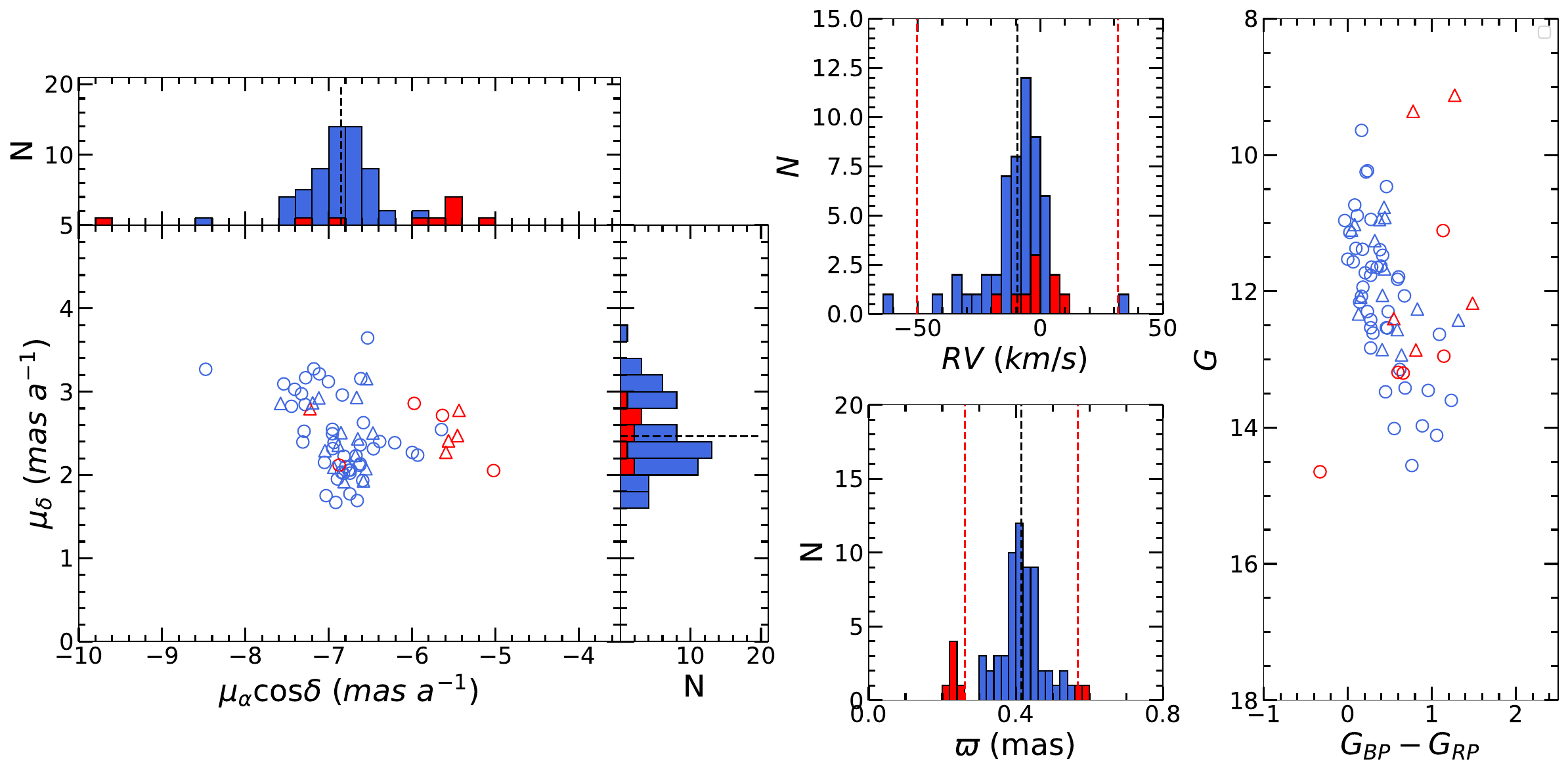}
\caption{Same as figure \ref{member_tr15} for the cluster Collinder 228}
\label{member_coll228}
\end{figure*}

\begin{figure*}
\centering
\includegraphics[width=\hsize]{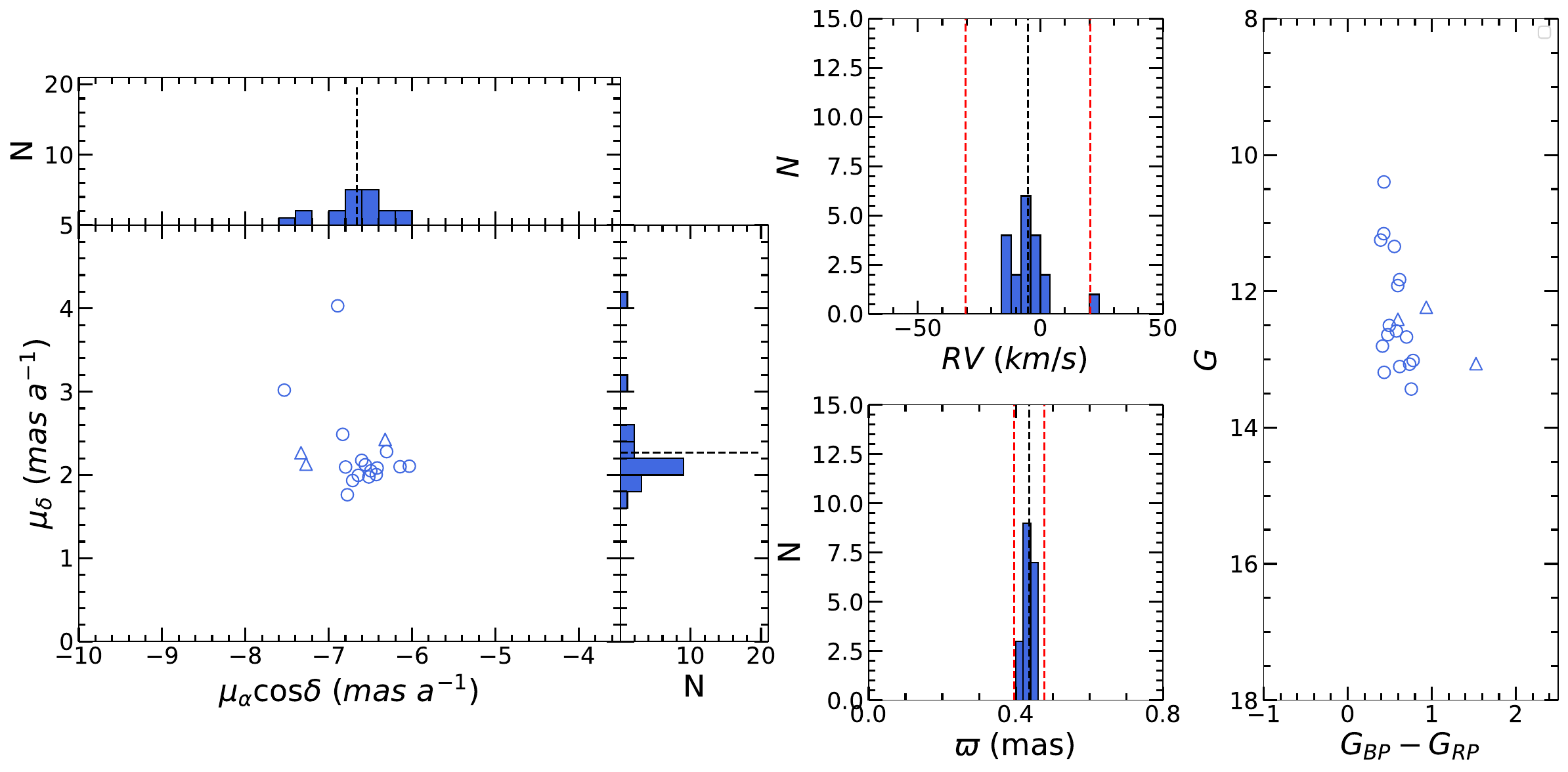}
\caption{Same as figure \ref{member_tr15} for the cluster Collinder 232}
\label{member_coll232} 
\end{figure*}

\begin{figure*}
\centering
\includegraphics[width=\hsize]{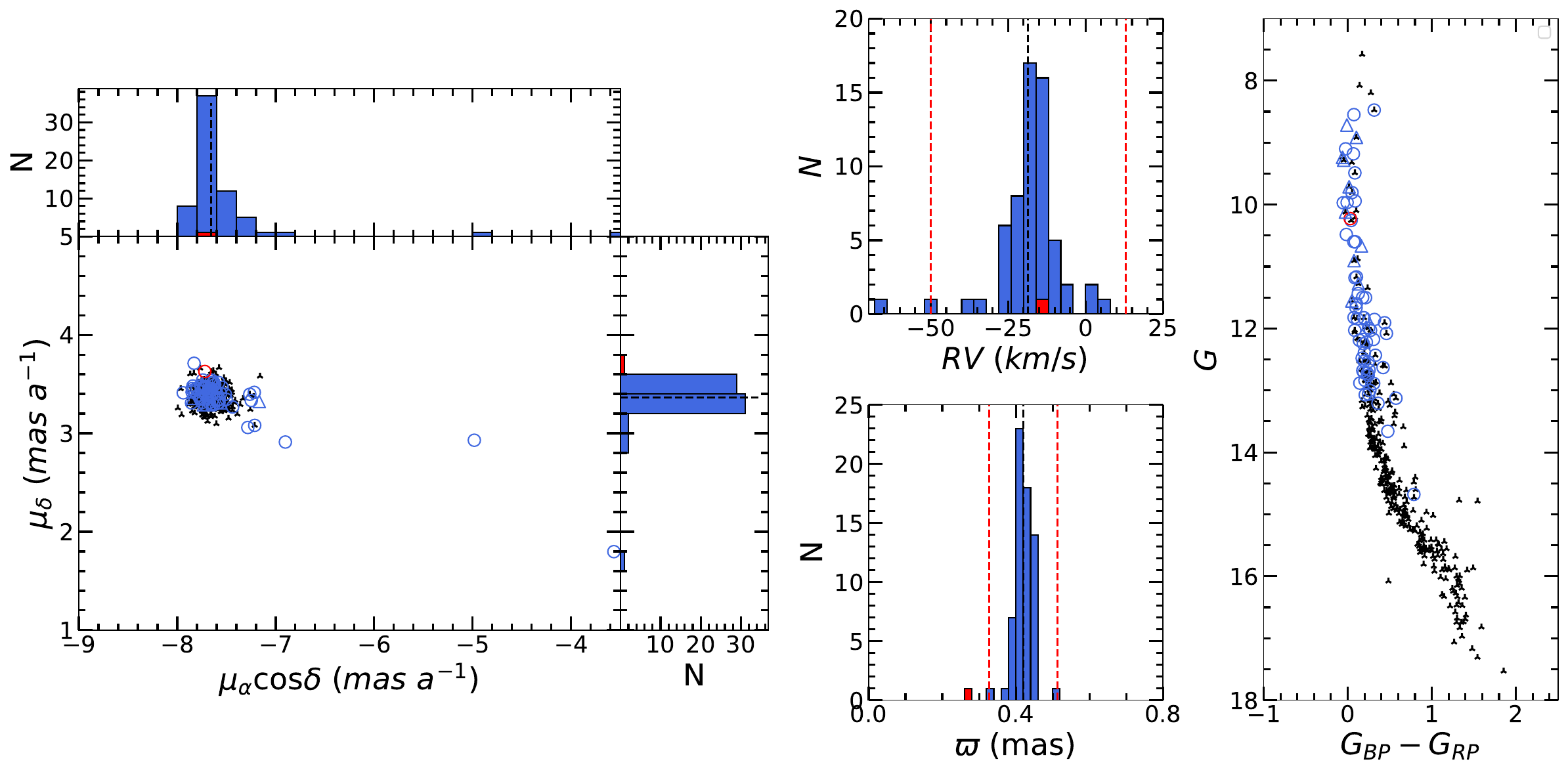}
\caption{Same as figure \ref{member_tr15} for the cluster NGC 3293}
\label{member_ngc3293}
\end{figure*}


\onecolumn

\section{Measurements of FHWM  and  estimates of projected rotational velocities }
\label{app:vsini} 

In this Appendix, we list the main results 
 for the stars analyzed in this paper: GES ID, others names, the radial velocities measured from He and Si lines, Signal-to-noise ratio (SNR) of the observed spectra,  full width at half maximum (FWHM (\AA)) measurements of the \ion{He}{i} lines 4388\AA\ and 4471\AA, the \Vsini\ estimates derived through the interpolation of the measured FWHM, and the average \Vsini\ values from \ion{He}{i} lines, with the corresponding dispersion. According to the analysis described in  Sec. \ref{member}, the stars are separated by cluster. 


 \onecolumn
 \pagebreak

\begin{longtable}{p{3cm}lccccccc} 
\caption{FHWM and \Vsini\ estimates for the sample stars \label{tab:vsini}}\\
\hline \hline
    GES ID	 & other ID	  & $V_{r}$ &	SNR	&	\multicolumn{2}{c}{FWHM  (\AA)} 	& \multicolumn{3}{c}{\Vsini\ (\kms)	}\\
   &   & (\kms) & & 4388\AA & 4471\AA	&  4388\AA	& 4471\AA	&	Average	\\
\hline
\endfirsthead
\caption{continued.}   \\
\hline\hline
GES ID & other ID	& $V_{r}$  & SNR	 &	\multicolumn{2}{c}{FWHM (\AA)} 	& \multicolumn{3}{c}{\Vsini\ (\kms)} 		\\
       &            & (\kms)   &         & 4388\AA     & 4471\AA	         &  4388\AA	& 4471\AA	&	Average	\\
\hline
\endhead
\hline
\endfoot
\hline
Trumpler 14 \\
\hline
10435522-5933147	& ALS 15 229	& $+2\pm$1 	&79	&	\dots	&	1.26	& \dots &	21	&	21	\\
10435847-5933016	& ALS 19 738	&	$-11\pm$3 	&175	&	5.96	&	5.99	&	247	&	223	&	235$\pm$17	\\
10435580-5932520	& [HSB2012] 1498 &	$-12\pm$1 	&104	&	3.47	&	4.52	&	135	&	148	&	141$\pm$9	\\
10435902-5933197	& 2MASS J10435902-5933196	&	$-21\pm$3 & 140 & 4.74 & 5.23 &	195	& 196 &	196$\pm$0 \\
10435650-5932498	& 2MASS J10435649-5932497	&	 $-7\pm$2  & 167 & 5.37 & 5.91 & 227	& 235 &	231$\pm$6	\\
10435366-5933006	& 2MASS J10435365-5933006	&	$-23\pm$2  & 144 & 2.86 &	3.32 & 112	& 111 & 111$\pm$0 \\
10435756-5933385	& ALS 15 224 &	$-10\pm$2 	&101	&	2.84	&	3.46	&	110	&	117	& 114$\pm$5	\\
10435724-5932412	& [HSB2012] 1550	&	 $-4\pm$2 	&86	& 	4.66	&	4.93	&	185	&	174	&	179$\pm$8	\\
10435952-5932316$^{e}$	& ALS 19739	 &	 $-8\pm$3 	&142	&	2.67	&	3.26	&	98	&	101	&	99$\pm$2	\\
10435208-5932401	& 2MASS J10435207-5932401	&	 $-3\pm$3  &47	& 3.56 & 4.46 & 136	& 146	& 141$\pm$7 \\
10435230-5932361$^{e}$        & 2MASS J10435230-5932360	&	  $+2\pm$1 & 110 & 3.46 & 3.93	& 129 &	129	& 129$\pm$0	\\
10435953-5932227	& 2MASS J10435953-5932227	&	 $+30\pm$1  & 93	& 1.88 & 2.57 &	67	& 73 & 70$\pm$4	\\
10435090-5933506	& ALS 19 733	&	 $-5\pm$2 	&171	&	4.92	&	5.69	&	205	&	219	&	212$\pm$10	\\
10440432-5932478$^{e}$	& 2MASS J10440432-5932478 	&	 $-1\pm$2  & 70	& 5.12 & 5.74 &	206	& 211 &	209$\pm$4	\\
10434798-5933591$^{e}$	& 2MASS J10434798-5933590	&	$-12\pm$2 & 83 & 5.35 &	4.84 & 219 & 219 & 219$\pm$0\\
10440063-5931524$^{e}$	& 2MASS J10440063-5931524	&	 $-5\pm$3 	&73	&	5.12 & 5.66	& 206 &	207	& 207$\pm$1	\\
10435603-5934410	& ALS 19 735 &	$-10\pm$1 	&137	&	3.06	&	3.77	&	117	&	123	&	120$\pm$4	\\
10440867-5933488$^{e}$	& 2MASS J10440866-5933488	&	 $-7\pm$3  & 95	& 8.13 & 8.60 & 339	& 350 & 344$\pm$8\\
10440384-5934344	& 2MASS J10440384-5934344	&	 $-4\pm$2  & 76 & 4.23 & 5.42 & 169 & 176 & 173$\pm$5	\\
10434478-5934158$^{e}$	& 2MASS J10434478-5934158	&	 $-2\pm$5 	&99	& 7.16 & 7.44 & 294 & 292 & 293$\pm$1	\\
10434356-5934035	& ALS 17 185 &	 $+ 4\pm$1 	&79	&	1.54	&	1.55	&	23	&	36	&	29$\pm$9	\\
10434580-5934359	& 2MASS J10434580-5934359	&	 $-6\pm$8  & 66	& 7.79 & 8.63 & 323	& 351 &	337$\pm$20	\\
10441215-5933512$^{e}$	& 2MASS J10441215-5933511	&	 $-1\pm$0  & 102	&	5.91 & 6.33	& 245 &	240	& 242$\pm$3	\\
10441283-5933322$^{e}$	& 2MASS J10441282-5933321	&	 $-6\pm$2  & 50	& 4.15 & 4.82 & 162 & 168	& 165$\pm$4	\\
10440583-5935117$^{e}$    	& ALS 15 227 &	$-12\pm$2 	&371	&	2.31	&	3.05	&	87	&	93	&	90$\pm$5 \\
10433597-5933179    	& 2MASS J10433596-5933179	&	 $-8\pm$1  & 80	& 1.90 & 1.79 &	43	& 45 & 44$\pm$1	\\
10433866-5934444	& 2MASS J10433865-5934444	&	 $-5\pm$1  & 106 & 8.86 & 9.29 & 377	& 388 & 382$\pm$8	\\
10435502-5936242	& 2MASS J10435501-5936242	&	$-10\pm$2 	& 81 & 3.80 & 4.79 & 157 &	161	& 159$\pm$3	\\
10440684-5936116	& 2MASS J10440683-5936116	&	 $-8\pm$1 	&82	& 7.59 & 7.81 &	313	& 311 &	312$\pm$2	\\
10435131-5928598	& Cl* Trumpler 14 Y 586		&	$-15\pm$2 	&58	& \dots & 2.64 & \dots &	94 & 94	 \\
10433085-5929239    & HD 303 312	 &	  $+9\pm$3 	&223	&	4.05	& 4.55 &	170	&	172	&	171$\pm$1	\\
10431695-5937534   & 2MASS J10431694-5937534	&	 $-9\pm$1 & 67	& 1.86	& 1.85	& 60 &	56	& 58$\pm$2	\\
10432367-5925595 & \dots & $-39\pm$1  & 55	& 2.17	&	2.10 & 65 &	66	& 65$\pm$1	\\
10432183-5924227$^{e}$	& ALS 16 078 &	$-28\pm$4 	&68	&	5.33	&	5.91	&	217	&	220	&	219$\pm$2	\\
10440093-5935459	& CPD-58 2625 &	 $-8\pm$1 	&163	&	1.39	&	1.51	&	23	&	38	&	30$\pm$11	\\
10441536-5936043$^{e}$	& 2MASS J10441535-5936042	&	 $-3\pm$0 & 88 & 5.74 & 6.55 &	237	& 250 & 244$\pm$10	\\
10435540-5926553	& 2MASS J10435539-5926551	&	 $-2\pm$3 & 45 &	3.50 & 3.88	& 131 &	130	& 131$\pm$1	\\
10440248-5929368	& ALS 1827	&	 $-2\pm$1 	&204	&	5.89	&	6.19	&	250	&	251	&	251$\pm$0	\\
10440828-5929594	& ALS 15 233 &	 $-8\pm$2 	&121	&	2.56	&	3.24	&	95	&	103	&	99$\pm$6	\\
10440509-5933413	& ALS 19 740 &	 $-6\pm$2 	&125	&	2.94	&	3.65	&	103	&	111	&	107$\pm$5	\\
10435177-5935451	& 2MASS J10435176-5935450	&	 $-7\pm$3 &40	& \dots & 7.85	& \dots	& 309 &	309	 \\
10433335-5935111	& ALS 1808	&	 $-6\pm$2 	&298	&	1.49	&	1.60	&	18	&	39	&	28$\pm$15	\\
10434124-5935530	& ALS 15203	&	$ -17\pm$2 	&49	&	3.44	&	3.96	&	140	&	139	&	139$\pm$1	\\
10434743-5931265$^{e}$	& 2MASS J10434743-5931264	&	 $-6\pm$1  & 104 & 1.53 & 1.11 &	30	& 23 & 26$\pm$5	\\
10440576-5927079	& 2MASS J10440576-5927078	&	  $+6\pm$0  & 51	& 1.96 & 1.92 & 50 & 53 & 52$\pm$2	\\
10435796-5933537$^{b}$	& ALS 15 864 &	 $-4\pm$2 	&146	&	2.18	&	2.38	&	74	&	72	&73$\pm$2 \\
\hline
Trumpler 15 \\
\hline
10444236-5922029	& 2MASS J10444235-5922029	&	 $-3\pm$1  & 68	& 4.63 & 5.10 &	186	& 180 & 183$\pm$4 \\	
10444212-5922305	& CPD-58 2655	&	 $-9\pm$2 	&144	&	4.22	&	4.85	&	175	& 179	& 177$\pm$3	\\	
10444278-5921383	&2MASS J10444278-5921383	&	        $+25\pm$4  & 44	& 5.25 & 5.71 & 213	& 210 &	212$\pm$2 \\	
10444178-5921330	& ALS 15856	&	$-11\pm$1 	&116	&	2.86	&	3.45	&	104	&	105	&	104$\pm$0	\\
10444946-5922192	& 2MASS J10444948-5922191	&	 $-9\pm$5  & 96	& 6.95 & 7.26 & 286	& 284 & 285$\pm$1 \\
10444058-5921138	& ALS 15 855 &	 $-5\pm$2 	&124	&	4.87	&	5.62	&	202	&	206	&	204$\pm$3	\\
10443514-5921220	& Cl Trumpler 15 24		&	  0$\pm$0 	&75	&	1.71	&	1.57	&	26 & 38	& 32$\pm$8	\\
10443511-5923282	& 2MASS J10443510-5923281	&	 $-7\pm$0  & 95	& 2.00 & 1.93 & 55	& 54 & 55$\pm$1	\\
10443075-5921263	& ALS 15 861&	 $-2\pm$1 	&149	&	1.53	&	1.57	&	18	&	36	&	27$\pm$12	\\
10443592-5923356	& 2MASS J10443591-5923356	&	 $-6\pm$2 	&92	& 5.54 & 6.25 &	227	& 236 &	232$\pm$7 \\
10443267-5920385	& Cl Trumpler 15 25		&	 $-8\pm$2 	&98	&	2.20 & 2.29 & 68 & 68 & 68$\pm$0	\\
10445073-5920234	& 2MASS J10445073-5920233	&	 $-9\pm$0  & 100	& 7.69 & 7.54 &	314	& 297 &	305$\pm$12	\\
10445869-5922228	& 2MASS J10445869-5922228	&	 $-3\pm$4  & 65	& 4.85 & 5.14 & 193	& 183 & 188$\pm$7	\\
10445733-5920476	& 2MASS J10445733-5920476	&	 $-5\pm$1  & 84	& 2.89 & 3.17 & 102	& 104	& 103$\pm$2	\\
10450531-5919349	& 2MASS J10450531-5919347	&	 $-2\pm$2  & 103 & 4.53 & 5.22 & 182 & 186 & 184$\pm$3	\\
10445081-5918004	& 2MASS J10445080-5918005	&	 $-7\pm$1  & 53 & 3.49 &	3.94 & 131	& 123 &	127$\pm$6 \\
10444209-592635$^{e}$	& 2MASS J10444208-5926353	&	 $-3\pm$0  & 79	& 8.14 & 8.70 & 336	& 353 & 344$\pm$12 \\
10440517-5921165	& 2MASS J10440516-5921165	&	 $-2\pm$1  & 63	& 4.86 & 5.18 & 197	& 189	& 193$\pm$6	\\
10440328-5919498	& 2MASS J10440327-5919498	&	  $+6\pm$1  & 74	& 1.99 & 2.16 & 52	& 58	& 55$\pm$3	\\
10441729-5917154	& 2MASS J10441729-5917154	&	 $-4\pm$1  & 68	& 3.38 & 3.67 &	125	& 117	& 121$\pm$6	\\
10440793-5917271	& 2MASS J10440789-5917267	&	 $-8\pm$1  & 54	& 3.27 & 3.74 & 120	& 121 &	120$\pm$0 \\
10452045-5917062	& HD 303300		&	$-16\pm$1 	&156	&	3.81	&	4.62	&	158	&	173	&	166$\pm$11	\\
10435355-5918466	& 2MASS J10435347-5918457	&	 $-8\pm$3  & 30 & 6.14 & 6.34 & 255 & 240 & 248$\pm$10 \\
10435416-5918244	& 2MASS J10435413-5918244	&	 $-6\pm$1  & 72	& 5.30 & 5.99 & 216	& 224 & 220$\pm$6	\\
10444398-5913590	& \dots &	$+ 18\pm$2 	&49	&	1.21	&	1.33	&	5	&	29	&	17$\pm$17	\\
10454824-5922042	& 2MASS J10454824-5922041	&	$-21\pm$2  & 63	& 6.32 & 6.87	& 261 &	267	& 264$\pm$4	\\
10435328-5916011	&	\dots &	  $+3\pm$6 	&58	&	4.27	&	4.63	&	167	&	161	&	164$\pm$4	\\
10440062-5925493	& ALS 1822	&	$-49\pm$1 	&216	&	2.16	&	2.27	&	76	&	72	&	74$\pm$3	\\
10443022-5926130	& TYC 8626-2506-1 &	 $-2\pm$2 	&128	&	4.61	&	5.01	&	196	&	194	&	195$\pm$1 \\
10432015-5917582	& 2MASS J10432014-5917582	&	$-22\pm$1 & 59	& 1.97 & 1.83 & 51 & 49	&	50$\pm$1 \\
10432557-5919175	& 2MASS J10432556-5919175	&	$-15\pm$1 	&59	& 3.29 & 3.99 & 121	& 123 & 122$\pm$1	\\
10444234-5923038	& ALS 1855	&	$-11\pm$1 	&166	&	3.41	&	4.15	&	140	&	152	&	146$\pm$9	\\
10444448-5921327	& ALS 15 854 &	 $-5\pm$2 	&221	&	5.57	&	6.31	&	234	&	237	&	235$\pm$2	\\
10442912-5920049	& ALS 1842	&	 $-2\pm$1 	&126	&	1.37	&	1.43	&	4	&	28	&	16$\pm$18	\\
10443766-5923073	& 2MASS J10443766-5923073	&	$ -1\pm$0  & 45	& \dots	& 6.85	& \dots & 265 &	265	 \\
10444065-5922285	& ALS 15 859 &	$ -4\pm$1 	&105	&	3.71	&	4.42	&	149	&	153	&	151$\pm$3	\\
10444652-5921538	& CPD -58 2662	&	 $-5\pm$1 	&76	&	2.65	&	3.28	&	91	&	97	&	94$\pm$4	\\
10441935-5917026	& GES J10441935-5917026	&	 $ -8\pm$3 	&62	& \dots	&	2.03	&	\dots	&	50	&	50		\\
10443636-5924203$^{b}$	& ALS 15 860	& $+5\pm$0 	&73	&	1.90	&	2.06	&	49	&	57	&	53$\pm$5	\\
10451811-5924277$^{b}$	& HD 93 342	&	 $+ 3\pm$1 	&75	&	1.40	&	1.07	&	23	&	21	&	22$\pm$2	\\
\hline
Trumpler~16~W \\
\hline
10443049-5941406$^{e}$	& ALS 15 215	&	 $-9\pm$2 	&141	&	5.36	&	5.83	&	227	& 232 &	230$\pm$3	\\
10442897-5942343$^{e}$	& ALS 19 742	&	$-10\pm$2 	&146	&	3.03	&	3.83	&	110	&	118	&	114$\pm$6	\\
10443139-5944080$^{e}$	& 2MASS J10443139-5944080	&	 $-9\pm$1  & 110 & 2.96 & 3.82 & 105 &	115	& 110$\pm$7	\\
10442895-5943473$^{e}$	& 2MASS J10442894-5943473	&	 $-9\pm$1  & 103	& 6.34 & 6.91 & 262 & 268 & 265$\pm$5	\\
10442377-5941065$^{e}$	& 2MASS J10442376-5941064	&	$ -4\pm$1  & 82 & 1.80 & 1.40 & 44 & 39 &	42$\pm$4 \\
10441969-5943079$^{e}$	& 2MASS J10441968-5943078	&	 $-1\pm$4  & 46	& 4.32 & 4.95 & 169 & 175	& 172$\pm$4	\\
10442251-5939258	& ALS 15 230	&	$-12\pm$1 	&165	&	6.86	&	7.43	&	294	&	299	&	296$\pm$4	\\
10441320-5943103	& ALS 15 210	&	$ -8\pm$3	&173	&	3.06	&	2.97	&	115	& 110 &	113$\pm$4	\\
10443290-5940261	& ALS 15 211	&	 $-9\pm$4 	&203	&	5.30	&	6.12	&	220	& 238 &	229$\pm$13	\\
\hline
Trumpler~16~E \\
\hline
10450180-5942014$^{e}$	& [HSB2012] 3314 &	 $-5\pm$3	&53	&	5.46	&	5.92	&	223	&	220	&	222$\pm$2	\\
10450216-5942010$^{e}$	& ALS 15 220	&	$-10\pm$2	&124	&	2.65	&	3.29	&	99	&	106	&	103$\pm$5	\\
10450523-5941426$^{e}$	& ALS 19 746	&	$-10\pm$1	&94	&	2.51	&	3.16	&	88	&	92	&	90$\pm$3	\\
10450673-5941565	& ALS 15200	 &	 $-6\pm$0	&228	&	1.36	&	1.44	&	20	&	33	&	27$\pm$9	\\
10450636-5942357	& ALS 15249	 &	 $-1\pm$1	&135	&	2.16	&	2.45	&	73	&	75	&	74$\pm$1	\\
10445478-5941239$^{e}$	& [HSB2012] 3100		&	 $-7\pm$4	&56	&	3.42	&	3.70	& 127 &	123	&	125$\pm$3	\\
10445408-5941294	& ALS 15 216	&	 $-7\pm$3	&177	&	7.09	&	7.81	&	300	&	321	&	311$\pm$15	\\
10450790-5941341$^{e}$	& HSB2012 3482	 &	$ -6\pm$0	&40	&	8.90	&	9.21	&	376	&	381	&	378$\pm$4	\\
10445796-5941031$^{e}$	& [HSB2012] 3192	&	$-23\pm$1	&119	&	5.12 & 5.66 & 208	&	208	&	208$\pm$0	\\
10450990-5942139$^{e}$	& HSB2012 3545	&	 $-9\pm$2	&90	&	7.26	&	7.91	&	302	&	318	&	310$\pm$12	\\
10450977-5942192	& HSB2012 3540	&	$-41\pm$2	&94	&	3.31	&	3.92	&	122	&	127	&	124$\pm$3	\\
10450520-5940574	& [HSB2012] 3397 &	 $-7\pm$1	&100	&	6.06	&	6.49	&	255	&	259	&	257$\pm$3 \\
10450933-5941283$^{e}$	& HSB2012 3526	&	 $-7\pm$2	&55	&	4.11	&	4.79	&	162	&	165	&	163$\pm$2	\\
10445991-5943149	& ALS 19744	 &	 $-5\pm$1 	&149	&	5.63	&	6.35	&	236	&	251	&	244$\pm$11	\\
10450590-5940546	& HSB2012 3424	&	 $-7\pm$1	&138	&	2.12	&	2.45	&	65	&	69	&	67$\pm$3	\\
10450974-5942572$^{e}$	& ALS 15246	 &	 $-9\pm$2	&122	&	2.72	&	3.29	&	102	&	104	&	103$\pm$2	\\
10451120-5941113	& ALS 15209	 &	 $-6\pm$3	&223	&	5.33	&	6.15	&	225	&	245	&	235$\pm$14	\\
10451170-5941163	& ALS 15213	&	 $-8\pm$0	&60	&	6.07	&	6.58	&	252	&	252	&	252$\pm$0	\\
10450023-5943345$^{e}$	& ALS 19 745	&	$-10\pm$1	&133	&	3.58	&	4.31	&	136	&	140	&	138$\pm$3	\\
10451400-5941420$^{e}$	& HSB2012 3648	&	 $-7\pm$1	&34	&	7.70	&	7.91	&	315	&	314	&	314$\pm$0	\\
10451265-5942488	& ALS 15228	&	$ -13\pm$1	&133	&	2.06	&	2.43	&	67	&	73	&	70$\pm$5	\\
10450020-5940052$^{e}$	& [HSB2012] 3263 &	$ -3\pm$0	&37	&	8.48	&	9.34	&	345	&	382	&	366$\pm$23	\\
10451622-5941412$^{e}$	& 2MASS J10451622-5941411 	&	$-12\pm$5 & 65	& 4.72 & 5.20 & 187	& 186 & 187$\pm$1	\\
10450968-5940088$^{e}$	& ALS 19747	 &	$-10\pm$2	&117	&	5.40	&	6.05	&	220	& 227	&	224$\pm$5	\\
10451355-5943318$^{e}$	& 2MASS J10451355-5943318	&	 $+ 7\pm$1	&34	&\dots	&	8.47 &\dots	& 341 &	341	\\
10451499-5943233	& ALS 15245		&	$-15\pm$1	&186	&	3.69	&	4.40	&	151	&	159	&	155$\pm$5	\\
10445846-5939437$^{e}$	& [HSB2012] 3211	&	  0$\pm$1	&27	&	2.33	&	2.74	&	78	&	77	&	78$\pm$1	\\
10451894-5942184$^{e}$	& 2MASS J10451894-5942184	&	 $-5\pm$1 & 62	& 1.69 & 1.77 &	23	& 43 & 33$\pm$14	\\
10451670-5943141	& ALS 15247		&	 $-5\pm$1	&96	&	2.95	&	3.65	&	112	&	118	&	115$\pm$5	\\
10450588-5944189$^{e}$	& ALS 15238	 &	 $-6\pm$2	&145	&	4.65	&	5.26	&	190	&	191	&	190$\pm$1	\\
10452057-5942213	& 2MASS J10452056-5942212	&	$ -8\pm$1  & 46 & 7.65 & 7.35 & 315 & 329 & 322$\pm$10	\\
10452057-5942513	& ALS 1878	&	 $-6\pm$1	&228	&	3.01	&	3.50	&	119	&	120	&	120$\pm$0	\\
10451656-5939571$^{e}$	& ALS 15223	&	$-12\pm$2	&130	&	5.10	&	6.04	&	216	&	236	&	226$\pm$14	\\
10450310-5944568$^{e}$	& [HSB2012] 3344 &	$ -4\pm$0	&29	&	\dots	&	7.11	& \dots	&	277	&	277	 \\
10452416-5942314$^{e}$	& 2MASS J10452415-5942313	&	 $-7\pm$0 & 79 &	7.75 & 8.20	& 320 &	330	& 325$\pm$7	\\
10443822-5943056$^{e}$	& 2MASS J10443822-5943056	&	 $-6\pm$5  & 102 & 8.11 & 8.25 & 336	& 333 & 335$\pm$3 \\
10450793-5939012	& 2MASS J10450792-5939011	&	$-10\pm$2 & 106 & 5.62 & 6.05 & 230	&	227	& 229$\pm$2	\\
10451272-5939066	& 2MASS J10451271-5939066	&	 $-3\pm$0 & 79	& 3.33 & 3.82 & 123	&	119	& 121$\pm$3	\\
10451943-5939374$^{e}$	& ALS 19748	 &	 $-8\pm$1	&126	&	3.26	&	4.08	&	122	&	131	&	127$\pm$6	\\
10450837-5938475	& 2MASS J10450836-5938475	&	$-32\pm$2 & 45	& 3.98 & 4.57 & 157	& 151 & 154$\pm$5 \\
10452314-5940034	& 2MASS J10452314-5940033	&	 $+ 1\pm$0	&83	&	3.42 & 4.14 & 127 &	133	& 130$\pm$4	\\
10452846-5941555$^{e}$	& 2MASS J10452846-5941555	&	$-20\pm$1	&18	& \dots	& 3.73 & \dots	& 122 &	122	 \\
10445394-5945240	& 2MASS J10445393-5945240	&	 $-3\pm$1 & 62 &	2.38 &	2.67 &	78 & 81	& 79$\pm$2	\\
10452521-5940011$^{e}$	& 2MASS J10452520-5940011	&	 $-6\pm$2 & 88	& 7.80 & 7.40 & 304	& 297 & 301$\pm$5	\\
10453134-5941133	& 2MASS J10453134-5941133	&	 $-9\pm$1 & 103 & 3.26 & 3.98 & 122 & 126	& 124$\pm$3	\\  
10452311-5944458	& 2MASS J10452310-5944458	&	 $+10\pm$2 &95 & 3.85	& 4.49 & 150 &	148	&	149$\pm$2	\\
10453255-5942359	& 2MASS J10453254-5942359	&	 $-5\pm$0 & 55	& 8.26 & 8.28 & 343	& 334 & 339$\pm$6	\\
10450823-5946070$^{e}$	& CPD -59 2629	&	$-10\pm$2	&170	&	1.53	&	1.60	&	35	&	43	&	39$\pm$6	\\
10452190-5945249	& 2MASS J10452190-5945249	&	$ -3\pm$1	&66	& 2.22 & 2.41 & 66 &	68 & 67$\pm$2	\\
10451297-5946060	& 2MASS J10451297-5946059	&	$ -6\pm$1	&56	& 5.00	& 5.52 & 202 & 201 & 202$\pm$1	\\
10444609-5946057	& 2MASS J10444609-5946056	&	$ -6\pm$1  & 84	& 2.54 & 2.84 & 87 & 83	& 85$\pm$2	\\
10453820-5942157	& 2MASS J10453819-5942157	&	 $+ 9\pm$1	&24	&	\dots	&	8.94	& \dots	& 367 &	367	 \\
10445364-5946595	& 2MASS J10445363-5946595	&	$ -1\pm$1 	&76	&	1.33 & 1.10	& 13 & 16 &	15$\pm$2 \\
10452214-5937385$^{e}$	& ALS 15244	&	 $-6\pm$1	&191	&	1.50	&	1.50	&	17	&	37	&	27$\pm$14	\\
10453808-5944095	& 2MASS J10453807-5944095	&	$ -8\pm$2	&63	&	1.26 & 1.36 & 11 & 28	& 19$\pm$12	\\
10454365-5939540	& 2MASS J10454365-5939540	&	$ -2\pm$1	&91	&	1.40 & 1.28	& 21 & 18 &	20$\pm$2 \\
10453368-5947148	& 2MASS J10453367-5947147	&	$ -4\pm$1	&121 & 2.75 & 3.08 & 94 & 101	&	98$\pm$4 \\
10454061-5937042	& 2MASS J10454060-5937041	&	$ -2\pm$1	&71	&	6.37 & 6.93 & 263 & 270	& 266$\pm$5	\\
10452228-5950471	& [ARV2008] 206	&	 $+ 3\pm$0	&102	&	2.48	&	2.62	&	93	&	85	&	89$\pm$6	\\
10461505-5940192	& Cl* Trumpler~16~ MJ 676 	&	$-18\pm$1	&22	& \dots	& 4.25 & \dots & 139 &	139	 \\
10445351-5951541	& 2MASS J10445351-5951541	&	 $-7\pm$1 	&148 & 3.71 & 4.32 & 141 & 148 & 145$\pm$5 \\
10443719-5940015	& ALS 15 242	&	 $-8\pm$1 	&145	&	2.28	&	2.81	&	81	&	88	&	85$\pm$5	\\
10444098-5940104	& ALS 19 743	&	 $-5\pm$1 	&152	&	1.17	&	1.21	&	2	&	17	&	10$\pm$11	\\
10445376-5937483	& ALS 15 236	&	$-19\pm$2 	&142	&	1.62	&	1.69	&	9	&	37	&	23$\pm$20	\\
10463082-5944176	& 2MASS J10463081-5944175	&	$ -7\pm$2	&90	& 2.93 & 3.03 & 103	& 98 & 101$\pm$4	\\
10444711-5939202	& 2MASS J10444710-5939201	&	$ -6\pm$1  &71 & 6.36	& 6.88	& 262 &	267	& 265$\pm$4	\\
10450579-5945196	& ALS 1870	&	  $+6\pm$0	&327	&	5.94	&	6.25	&	253	&	254	&	254$\pm$0	\\
10450584-5943077	& ALS 15 197	&	$-35\pm$1	&200	&	1.48	&	1.57	&	31	&	41	&	36$\pm$8	\\
10453661-5944111	& 2MASS J10453660-5944110	&	 $+ 5\pm$2	& 46	& 6.00 & 6.32 & 256 & 257 & 257$\pm$1	\\
10445602-5938530	& 2MASS J10445602-5938530	&	 $-6\pm$2 & 45 &	4.49 &	5.26 & 181	& 189 &	185$\pm$5 \\
\hline
Collinder 228 \\
\hline
10443687-6001116	& ALS 15957	 	&	 $-5\pm$1 	&246	&	8.45	&	9.16	&	358	&	381	&	370$\pm$17	\\
10442198-5959351	&2MASS J10442207-5959351 	&	 $-5\pm$1  & 138 & 3.74 & 4.62	& 152 &	150	& 151$\pm$1	\\
10441496-6000057	& ALS 1836	&	+34$\pm$2 	&499	&	2.30	&	2.56	&	84	&	79	&	81$\pm$4	\\
10441444-6001270	& ALS 15 958 	&	$-10\pm$1 	&144	&	3.69	&	4.52	&	145	&	153	&	149$\pm$6	\\
10443829-6005450	&2MASS J10443829-6005449 	&	$-60\pm$2 	&84	& 3.57 & 4.17 &	135	& 135 &	135$\pm$0 \\
10443174-6005449$^{e}$	& ALS 16 054 &  $-6\pm$1 	&139	&	4.62	&	5.23	&	189	&	191	&	190$\pm$1	\\
10441513-6007509	& CPD-59 2569B	&	 $-3\pm$1 	&169	&	1.86	&	1.99	&	44	&	53	&	48$\pm$7	\\
10440078-6006012$^{e}$	& CPD-59 2554	&	$-16\pm$1 	&139	&	4.53	&	5.11	&	181	&	181	&	181$\pm$0	\\
10440104-6006378        & 2MASS J10440105-6006377	&	 $-2\pm$1 	& 95 & 2.13 & 2.19 & 63	& 64 & 64$\pm$1	\\
10435479-6006208$^{e}$    	& 2MASS10435478-6006207	 	&	 $-1\pm$2  & 95	& 2.41	& 2.84	& 79 & 86 &	83$\pm$5 \\
10435238-6001175$^{e}$	&	\dots	&	 +3$\pm$1 	&90	&	6.77	&	7.70	&	279	&	304	&	291$\pm$18	\\
10434797-6001201$^{e}$	& 2MASS J10434797-6001201	&  +2$\pm$1 	& 92	& 4.44 & 5.02 & 176 & 176 & 176$\pm$0	\\
10434887-6006437$^{e}$    	&2MASS J10434886-6006437 	& +6$\pm$1 	&126 &	5.11 &	5.74 & 206	& 212 &	209$\pm$4	\\
10434679-6008264$^{e}$	& CPD-59 2539	&	 +2$\pm$1 	&130	&	2.45	&	2.84	&	81	&	87	&	84$\pm$4	\\
10434887-6009009	& ALS 18775	&	$-26\pm$2 	&156	&	4.54	&	5.14	&	183	&	181	&	182$\pm$1	\\
10435198-6010368$^{e}$	& 2MASS J10435127-6010386	&	        $+3\pm$3  & 86	& 6.24 & 6.76 & 258 & 261 & 259$\pm$2	\\
10433156-6003160$^{e}$	& 2MASS J10433154-6003159	&	       $ -3\pm$1  & 254 & 3.01	& 3.44	& 107 &	109	& 108$\pm$1	\\	
10432030-6013015	& 2MASS J10432028-6013014	& $ -6\pm$1 	& 107 & 4.80 & 3.77 & 179 & 119 & 149$\pm$42 \\
10424533-6012063	& 2MASS J10424532-6012063	& $-32\pm$1 	& 144 & 2.51 & 3.15 &	86 & 91	& 88$\pm$4	\\
10425717-6007414   	& CPD -59 2510	& $-4\pm$1 	&182	&	1.75	&	1.74	&	20	&	38	&	29$\pm$13	\\
10425445-6002595	& 2MASS J10425444-6002594	& $-41\pm$2 	&96	&	2.13 & 1.90 & 62 &	62	& 62$\pm$0	\\
10425898-6000242	& 2MASS J10425900-6000240	&	$-11\pm$2  & 108 & 3.61 &	4.17 &	137	& 136 &	136$\pm$1 \\
10424616-6000576	& CPD-59 2504	&	$-3\pm$3 	&138	&	5.21	&	5.90	&	211	&	219	&	215$\pm$6	\\
10422722-6000052    	& 2MASS J10422722-6000052	&	$-16\pm$3 	&84	& 3.74 & 4.45 &	144	&	146	& 145$\pm$1	\\
10423616-5959262	& ALS 15956	&	$-12\pm$1 	&260	&	2.92	&	3.54	&	104	&	106	&	105$\pm$1	\\
10424804-5953371	& \dots	&	$-14\pm$2 	&131	&	4.27	&	4.61	&	167	&	159	&	163$\pm$6	\\
10421831-6001553	&	\dots 	&	$-14\pm$4 	&104	&	2.27	&	2.21	&	70	&	70	&	70$\pm$0        \\
10422123-5957156	&	\dots 	&	$-22\pm$2 	&177	&	3.84	&	4.08	&	148	&	136	&	142$\pm$8	\\
10420759-5956249	&	 \dots 	&	$-16\pm$1 	&140	&	1.68	&	1.38	&	38	&	38	&	38$\pm$0	\\
10415981-5955075	& HD 305439	 & $ -23\pm$1 	&159	&	3.01	&	3.72	&	107	&	115	&	111$\pm$6	\\
10434882-6000366$^{e}$	& ALS 15 962 	&	$-6\pm$2 	&114	&	4.86	&	6.01	&	204	&	228	&	216$\pm$17	\\
10440371-5948141	& 2MASS J10440371-5948141	& $-10\pm$1 	&93	&	2.21	&	2.74 & 76 &	83	& 79$\pm$5	\\
10434303-5945334	& 2MASS J10434303-5945333	& $-10\pm$1 	&70	& 3.61 & 4.31 & 138	& 137 &	138$\pm$0	\\
10434813-5950443	& Cl* Trumpler 14 MJ 137 	&	 $-3\pm$2 	&117 & 2.24	& 2.51	& 72 & 70 & 71$\pm$1	\\
10435010-5947025$^{e}$	& 2MASS J10435009-5947024	&	$-13\pm$2  &53	& 4.82 & 5.09 & 193	&	180	& 187$\pm$9	\\
10435088-5950308	& 2MASS J10435088-5950307	&	+10$\pm$4 	&59	& 4.08 & 4.41 &	158	&	151	&	155$\pm$6	\\
10440071-5949519	& 2MASS J10440071-5949518	&	$-3\pm$2	&128 & 1.94 & 1.64	& 52 & 51 &	52$\pm$1	\\
10440237-5952047	& 2MASS J10440236-5952046	&	$-17\pm$1	&108 &	1.69 & 1.49	& 23 &	33	& 28$\pm$7	\\
10441456-5954492$^{e}$	& 2MASS J10441455-5954492	&	$-17\pm$1 	&39	& \dots & 5.19 & \dots & 186	&	186	 \\
10441879-5951490$^{e}$	& 2MASS J10441879-5951490	&	$-32\pm$5 	&90	& 4.80 & 5.39 & 199	& 203	& 201$\pm$3	\\
10444512-5954114$^{e}$	& 2MASS J10444512-5954114	&	$-10\pm$2 	&44	& 2.96 & 3.05 & 105	& 104 &	105$\pm$1	\\
10445041-5955450	& ALS 1864		&	 $-2\pm$1 	&267	&	1.99	&	2.17	&	68	&	70	&	69$\pm$2	\\
10445392-5956134 	& Gaia DR3 5350302097055370496	&	$ -8\pm$2  & 39 &	4.66 & 5.27	& 188 &	188	& 188$\pm$0	\\
10450602-5956512	& 2MASS J10450601-5956512	&	$-8\pm$2 	&52	& 1.22 & 0.99 & 1 & 6 &	4$\pm$3	 \\
10451073-5957548	& 2MASS J10451072-5957547	& $-5\pm$1	&51	&	6.57 & 7.53	& 271 &	297	& 284$\pm$18	\\
10451338-5957538	& HD 305533	&	 $-5\pm$2	&214	&	4.69	&	5.49	&	199	&	217	&	208$\pm$13	\\
10451341-6000589	& CPD-59 2638	&	 $-9\pm$1	&105	&	2.74	&	3.11	&	95	&	95	&	95$\pm$0	\\
10451672-5954458	& HD 305528		&	 +2$\pm$0	&117	&	2.16	&	2.25	&	65	&	69	&	67$\pm$3	\\
10433443-5943265$^{e}$	& 2MASS J10433443-5943264	&	$-2\pm$1  & 92	& 1.42	& 1.47	& 26 & 35 &	30$\pm$7 \\
10443008-5952141	& ALS 15 222	&	 $-6\pm$1 	&178	&	1.13	&	1.22	&	4	&	15	&	10$\pm$8	\\
10445734-6000467	& ALS 1866	&	 $-5\pm$1	&157	&	2.06	&	2.21	&	68	&	68	&	68$\pm$1	\\
10445053-5957227	& 2MASS J10445053-5957227	&	 +1$\pm$0 	&46	& 2.37 & 2.87 & 81	& 82 &	81$\pm$1	\\
10444551-5952538	& 2MASS J10444550-5952537	&	$ -9\pm$3  & 30 &	5.48 & 5.59	& 226 &	209	& 217$\pm$12	\\
10435403-5946106	& 2MASS J10435403-5946105	&	$-12\pm$1  &21	& \dots	&	5.69 &	\dots &	193	& 193 \\
10413434-5958474$^{b}$$^{,e}$	&2MASS J10413434-5958474 	& 	$-20\pm$1 	&94	& 2.22	& 2.51	& 77 & 77 &	77$\pm$0 \\
10421033-5958009$^{b}$	& HD 305439		&	$-11\pm$2 	&282	&	2.06	& 2.41 & 69	& 75 & 72$\pm$4	\\
10425293-6003478$^{b}$$^{,e}$	& 2MASS J10425293-6003478	&	$ -3\pm$1  & 62	&	2.06 &	2.23 &	63	& 63 & 63$\pm$0	\\
10443676-5954249$^{b}$	& ALS 1853	&	 +6$\pm$1 	&105	&	1.61	&	1.76	&	7	&	41	&	24$\pm$24	\\
10442910-5948207$^{f}$	& 2MASS J10442909-5948207	& $-7\pm$0 	&60	& 1.98 & 2.22 &	67	& 72 &	69$\pm$3	\\
\hline
Collinder 232 \\
\hline
10444614-5933041$^{e}$	& 2MASS J10444614-5933041	&	$-6\pm$3  & 56	& 6.09	& 6.54	& 252 &	250	& 251$\pm$1	\\
10445838-5932063$^{e}$	& 2MASS J10445837-5932062	&	$-7\pm$1 & 67 & 6.67 & 7.36 & 275 &	290	& 283$\pm$10 \\
10445036-5934469	& 2MASS J10445036-5934468	&	$-12\pm$1  & 39	& \dots	& 3.60 & \dots	& 119 &	119	 \\
10452417-5932358	& 2MASS J10452417-5932357	&	 $-1\pm$1 &85	& 3.60 & 3.87 &	136	& 126 &	131$\pm$8	\\
10444734-5926595$^{e}$	& Gaia DR3 5350388683629610752	&	    $-6\pm$1 & 99 &	6.15 &	6.51 &	253	& 249 &	251$\pm$3 \\
10451589-5929564	& 2MASS J10451588-5929563	&	$-14\pm$4 	&78	& 7.34	& 7.67	& 308 &	310	& 309$\pm$2	\\
10451925-5929522$^{e}$	& 2MASS J10451925-5929522	&	        +2$\pm$0  & 79	& 1.36 & 1.47 &	8 &	32	&	20$\pm$17	\\
10452876-5930038	& 2MASS J10452875-5930037	&	 +3$\pm$1 	&19	& 1.18 & 1.28 &	6	&	16	&	11$\pm$7 \\
10442518-5928160	& ALS 15 232	&	 $-8\pm$2 	&89	&	4.72	&	5.33	&	204	&	204	&	204$\pm$0	\\
10442886-5928166	& 2MASS J10442885-5928165	&	$-2\pm$1  & 72	& 1.26	& 1.25	& 7	& 12 & 9$\pm$4   \\
10444727-5928155   	& 2MASS J10444727-5928155 	&	$-2\pm$1  & 93	& 4.65	& 5.19	& 186 &	184	& 185$\pm$1	\\
10450616-5931231	& ALS 15237	&	 +22$\pm$2 	&148	&	4.37	&	5.27	&	176	&	189	&	182$\pm$9	\\
10442549-5933093	& ALS 15 235 &	$-16\pm$2 	&110	&	2.87	&	3.68	&	104	&	114	&	109$\pm$7	\\
10442848-5932229	& 2MASS J10442848-5932228	&	 $-1\pm$1  &61 &	4.33 &	4.76 & 170	& 167 &	168$\pm$2 \\
10442946-5933438	& 2MASS J10442945-5933437	&	 $-6\pm$1 & 68 &	7.95 &	7.98 & 328	& 319 &	324$\pm$6	\\
10443223-5933593	& 2MASS J10443223-5933592	&	$-15\pm$1  &138	& 7.62 & 7.42 & 317 & 294 & 305$\pm$16 \\
10444650-5934134	&2MASS J10444649-5934134	&	 $-8\pm$1 & 72	& 4.97 & 5.42 &	199	& 196 &	198$\pm$2	\\
10445629-5933035 	& ALS 15 205	&	 $-7\pm$2 	&158	&	5.40	&	6.10	&	230	&	246	&	238$\pm$12	\\
10442598-5928593$^{e}$	& Gaia EDR3 5350387270553228928	&	$-11\pm$2 & 98	& 2.43	& 3.16	& 93 & 94 & 94$\pm$1 \\
\hline
Bochum 11 \\
\hline
10452563-5958452	& 2MASS J10452563-5958451	&	$-6\pm$2 & 27 &	4.74 &	5.46 & 188	& 198 &	193$\pm$7	\\
10453186-6000294	& AHP2016 OBc 53 &	$-27\pm$4	&24	&	\dots	&	3.99	& \dots	&	144	&	144		\\
10454701-6000272	& 2MASS J10454701-6000271	&	 $-5\pm$1 & 96 &	3.53 &	4.42 &	133	& 145 &	139$\pm$8	\\
10455900-5958098	& 2MASS J10455899-5958098	&	  +2$\pm$0  & 111 &	6.93 &	7.12 &	285	& 277 &	281$\pm$5	\\
10460116-5949420	& 2MASS J10460116-5949420	&	  +8$\pm$2	&13	& \dots	& 5.17	& \dots	&	187	&	187		\\
10460277-5950193	& 2MASS J10460277-5950192	&	$-10\pm$1 &75 & 7.89	& 8.48	& 330 &	347	& 338$\pm$12 \\
10460291-5950259	& 2MASS J10460291-5950259	&	 $-3\pm$1 & 20 &	2.05 &	2.05 &	58	& 53 &	56$\pm$4 \\
10460478-5949218	& 2MASS J10460477-5949217	&	+6$\pm$2	& 120 & 4.94 & 5.46 & 210	& 216 & 213$\pm$4	\\
10460493-5959018	&	\dots	&	  +6$\pm$0	&175	&	1.92	&	2.05	&	44	&	53	&	48$\pm$6	\\
10460607-5956339	& 2MASS J10460606-5956339	&	$-3\pm$1 & 123	& 4.78	& 5.33	& 193 &	191	& 192$\pm$1	\\
10460609-5957394	& 2MASS J10460608-5957394	&	 $-1\pm$1 & 24 & \dots &	2.69 & \dots &	82	&	82 \\
10462246-5953205	& ALS 1892	&	$-15\pm$1   &161	&	\dots	&	5.35	&	\dots	&	230	&	230		\\
10462658-5956131	& 2MASS J10462657-5956131	&	$-11\pm$0	&19	& \dots	& 4.10	& \dots	& 127 &	127	 \\
10463118-5954352	& 2MASS J10463117-5954351	&	$-25\pm$1	&17	& \dots	& 3.03	& \dots	&	96	&	96	\\
10463278-5952375	& 2MASS J10463277-5952375	&	$-18\pm$3	&8	& \dots	& 2.34 & \dots	&	68	&	68	\\
10463643-5948049	& 2MASS J10463643-5948048	& $-8\pm$1	&43	&	1.55	& 1.57 & 11	& 32	& 21$\pm$15	\\
10463801-5955166	& 2MASS J10463801-5955165	& $-8\pm$6	&48	&	6.11 & 6.63 &	253	&	255	& 254$\pm$1	\\
10464887-5950410	& 2MASS J10464886-5950409	&	$-10\pm$1	&57	&	4.34 & 5.08	& 176 &	183	& 179$\pm$4	\\
10470064-5957242	& 2MASS J10470063-5957242 &	$-8\pm$1	&56	& 3.15	& 3.95	&	114	&	123	&	118$\pm$7	\\
10470436-5948281	& 2MASS J10470435-5948281	& $-2\pm$1	&46	& 6.14 & 6.28 & 255	& 237	& 246$\pm$12	\\
10470657-5951515	& 2MASS J10470656-5951514	& +8$\pm$1	&45	&	2.27	& 2.21 & 70 & 71 & 71$\pm$1	\\
\hline
NGC 3293 \\
\hline
10342068-5814107	&  \dots &   +4$\pm$9	&77   & 8.4	&    4.2	&   134	& 134 & 	134$\pm$0	\\ 
10342078-5813305	& CPD-57 3450 	&   $-21\pm$2	& 176  	&    3.4	&    3.8 &   129	&  116	&   123$\pm$9	\\
10344202-5815419   & ALS 20075  &  $-26\pm$2	& 71   	&    3.5	&    3.1	&  129	&  102	&   115$\pm$19	\\ 
10344494-5809103$^{e}$	&  \dots &   $-16\pm$10 	& 7  &    1.5	&    1.7	&  	42	&   50	&   46$\pm$6	\\ 
10344774-5807274	& 2MASS J10344773-5807272 	&  $-22\pm$5 & 49  & 4.7 & 4.9	& 190 &  175 &  183$\pm$10	\\ 
10344868-5809012	&  V* V400 Car &  $-16\pm$1	&41   	&    1.9	&    2.2	&    	63	&   68	&  65$\pm$3	\\ 
10344869-5807127	&  \dots	& 	$-11\pm$2	&298  	&    2.8	&    3.1	&    	98	&   101	&  99$\pm$2	\\ 
10345480-5807233     & ALS 20077 &  +1$\pm$4  & 71   	&    5.4 &  5.4 &  220	&  194	&  207$\pm$18	\\ 
10350325-5814268    & ALS 20084 &   +1$\pm$4 	&72  	&    6.6	&    6.1	&   273	&  228	&  251$\pm$31	\\ 
10351355-5811124	& 2MASS 10351352-5811125 &  $-38\pm$4 	&89  & 4.5	&  5.5	&  181	&  202	&  192$\pm$15	\\ 
10352272-5817081    & ALS 20080 &  	$-19\pm$4 	&78   	&    6.9	& 7.1 &  283	&  275	&  279$\pm$6	\\ 
10352410-5813284 &  2MASS J10352408-5813285 	&  $-14\pm$3 	&103 & 4.7	& 5.2 & 190	& 186 & 188$\pm$2	\\ 
10352851-5812496	& 2MASS J10352849-5812497 	&   $-52\pm$1  &224  & 1.5	&  1.2	& 27 & 5 &	16$\pm$16	\\ 
10353007-5812080 	& HD 303067 &  $-14\pm$2   &291    &    2.8	&    3		&    	103	&   92	&  97$\pm$8	\\ 
10353230-5815220$^{e}$ 	& HD 303075 &  $-13\pm$5 &207  	&    6.6	&    6.6	&    	281	&  270	& 275$\pm$8	\\ 
10353289-5810162 & Cl* NGC 3293 FEAS 454 & $-16\pm$2 	&51  	&    5.7 & 3.9 & 131	& 131	&   131$\pm$0	\\ 
10353568-5813564	& 2MASS J10353565-5813566  &  $-15\pm$5 	&76 &  3.4 & 4.4 & 128	& 145	&   136$\pm$12	\\ 
10353581-5813077 & Cl* NGC 3293 ESL 116  & $-15\pm$14   	&84	&  7.5 & 6.5 &  307	& 246	&  276$\pm$43	\\ 
10353662-5816040 &  2MASS J10353660-5816042 &  $-20\pm$2 	&108  & 4.1 &  4.8 & 	163	&  166	&   164$\pm$2	\\ 
10353767-5812120 &  2MASS J10353767-5812120 	& $-9\pm$3   	&148 & 6.7	& 6.3	& 277	& 240	&  259$\pm$26	\\ 
10353848-5813069	& Cl* NGC 3293 ESL 73 	& $-19\pm$3 	&209  &  4.2 & 3.7 & 164	& 124	&   144$\pm$28	\\ 
10354026-5813012 & 2MASS J10354026-5813012 	&  $-5\pm$7   	&133 & 7 & 6.1 & 287	&  227	&    257$\pm$42	\\ 
10354072-5812440    	& ALS 1671 	&  $-22\pm$1 	&64  	&   1.6 & 2.2 &  51	&  70	&  60$\pm$13	\\ 
10354194-5811567 & 2MASS J10354191-5811568 	&  $-19\pm$5 	& 550 & 7.3	& 7.4	& 308	& 300	&  304$\pm$6 \\ 
10354331-5813334    	& ALS 20065 &  $-25\pm$3 	&153  	&    5 &   5.5 &  210	&  210	&  210$\pm$0	\\ 
10354405-5813459 & Cl* NGC 3293 ESL 40  & $-24\pm$4 	&129  &  7.9	&  7.4	&  330	& 291	& 310$\pm$28	\\ 
10354456-5814303 	& V438 Car & $-19\pm$1 	&275 &  4.8	& 5.3	& 194	&  193	&	193$\pm$16	\\ 
10354499-5816350 &  2MASS J10354497-5816352	& $-16\pm$2 	&217  & 4.5	& 5.5	& 179	& 201	& 190$\pm$16	\\ 
10354530-5815280 &  Cl* NGC 3293 ESL 57 	& $-17\pm$3 	&98  	&    5.5 & 6.1	&  226	&  228	&  227$\pm$2 \\ 
10354723-5813375 	& ALS 15740 & $-17\pm$2	&174   	& 2		&    1.6 &  56	& 31	&	44$\pm$18	\\ 
10354752-5805130 	& ALS 20081 &  $-18\pm$7 	&117   	&    6.5	&    6.6 &   270	&  252	&  261$\pm$13	\\ 
10354822-5812329  & ALS 15741 & $-17\pm$1	&68 	&    2.8	&    3.5	&    	108	&    	119	& 114$\pm$8	\\ 
10354901-5814541    	& ALS 20070 & $-18\pm$4 	&313    &    2		&    3.5 &  62	&  111	&  87$\pm$34	\\ 
10354954-5815398 & 2MASS J10354954-5815398 	& $-12\pm$2 	&221   	&    5.8 & 6.5	& 237	&  248	& 242$\pm$8	\\ 
10355051-5812126	&  2MASS J10355050-5812127	& $-26\pm$2 	&73	& 6.4 &  5.5 & 208	& 208	& 208$\pm$0	\\ 
10355073-5811414 &  2MASS J10355073-5811414 & $-16\pm$3 	& 194  &  5.4 & 5.5 & 221	& 200	&  211$\pm$14	\\ 
10355129-5816179    	& ALS 20087 &  $-18\pm$4 & 105    &    6.2	&    6.4	& 257	&  247	&  252$\pm$8	\\ 
10355174-5813392 &  2MASS J10355174-5813392	& $-16\pm$6  	&106    &  6 & 5.6 &  246	& 205	&  225$\pm$28	\\ 
10355301-5812168 	&  ALS 15743 & $-25\pm$2  	&271    &    5.1	&    5.5	&  211	&  212	&  211$\pm$1	\\ 
10355312-5814260 &  2MASS J10355311-5814257	& $-27\pm$6 	&53  &  5.6 &  4.6 &  235	& 164	& 200$\pm$50	\\ 
10355363-5814478 	& ALS 15744 & $-18\pm$18 	&70  & 4	& 5.1 &  167	&	200	&  183$\pm$23	\\ 
10355397-5815047   	& ALS 20092  & $-11\pm$5 	&94    	&    5.6	&    3.7	&    	121	&  121	&  121$\pm$0 \\ 
10355422-5815267 	& HD 91983	&  	$-25\pm$2 	&101 	&    3.1	&    3.8	&  122	&  135	&  129$\pm$9	\\ 
10355467-5813486$^{e}$ 	& ALS 15745 & $-18\pm$3  & 125  	&    8.8	&    8.1	&	370	& 326	& 348$\pm$31	\\ 
10355469-5812371 & 2MASS 10355469-5812371  	& $-13\pm$2 	& 195 & 6.2 & 5.8 &  257	&  217	&  237$\pm$28	\\ 
10355491-5812591  	& ALS 20064 & $-18\pm$3   	&77  	&    6.6	&    7.1 &  276	&	289	&  282$\pm$9	\\ 
10355492-5806492	& 10355492-5806492 		& $-15\pm$3 	&152   	&    4.5 &  4 &  179	&  137	& 158$\pm$30	\\ 
10355539-5812197   & ALS 20066 & $-18\pm$1 	&237  	&    4		&    4.8 &  161	& 177	&  169$\pm$11	\\ 
10355660-5811314  & ALS 15746 &  $-22\pm$3 	&264   	&    1.5	&    1.4	&    	34	&   21	& 	27$\pm$10	\\ 
10355711-5815218	& Cl* NGC 3293 ESL 66 &  $-35\pm$1   	&126   	&    1.9 &  1.2 &  48	&  27	&  38$\pm$15	\\ 
10355781-5812213  & ALS 15748 &  $-15\pm$2	&375  	&    1.3	&    1.1	&    	21	&  5	&  13$\pm$11	\\ 
10355836-5814411 & 2MASS J10355836-5814411 	&  $-25\pm$3  & 44  &  5.6 & 5.2 & 230	& 184	& 207$\pm$32 \\ 
10355849-5814148  	& ALS 15751 &  $-28\pm$8 	& 77  	&    5.2 & 5.9 &  218	& 228	& 223$\pm$7	\\ 
10360160-5815096 	& HD 92007 & $-21\pm$1  &370   	&    2.2	&    2.6	&    	75	& 77	& 76$\pm$1	\\ 
10360229-5812581 & 2MASS J10360231-5812583  	& $-19\pm$2 	&73 & 6.3 &  6.2 &  259	&  234	& 246$\pm$18	\\ 
10360349-5814401 	& ALS 15753 & $-15\pm$1  	&198  &    5.4	&    6		&    	221	&    	223	& 222$\pm$1	\\ 
10360491-5810433 	& HD 303065 & $-13\pm$2 	&244  	&    3.4	&    3.8	&    	139	& 135	& 137$\pm$2	\\ 
10360525-5816455 & 2MASS J10360523-5816456  	& $-66\pm$2 	&63  &  1.4 &  2.5 &  22	&  68	&  45$\pm$32	\\ 
10360595-5814270$^{e}$ 	& ALS 1683 & $-20\pm$3  	&219   	&    8.5	&    7.8	&  358	&  313	&  335$\pm$32	\\ 
10360656-5806551 & 2MASS J10360656-5806549 	&  $-19\pm$2 	& 67 &  5.7 &    5.5 & 233	& 198	&  216$\pm$25	\\ 
10360657-5817538 	& ALS 20074 & $-10\pm$1 	&83  	&    7.4	&    6.6	&   301	&  255	& 278$\pm$33	\\ 
10360764-5815204 	& ALS 20072 & $-18\pm$2 & 204 &    2.3	&    2.8	&    	78	&  85	&  81$\pm$5	\\ 
10361290-5813250	& 2MASS J10361290-5813250  & $-7\pm$1 	&106	& 6.8 & 6.2 &  281	& 237	&  259$\pm$31	\\ 
10361348-5811207 & 2MASS J10361347-5811208  &  $-24\pm$1 	&90   	&    4.5 & 4.8 & 181	& 165	& 173$\pm$12	\\ 
10361692-5815000 	& ALS 20085 & $-22\pm$2 	&53   	&    6.5	&    5.4	&  197	& 197	&  197$\pm$0	\\ 
10362800-5808180	& 2MASS J10362800-5808180 &  $-16\pm$2 	& 61  &  7.1 & 6.9 &	290	&  268	& 279$\pm$15	\\
\hline
\hline
{\footnotesize$^f$: probable foreground star kept in the sample}\\
{\footnotesize$^b$: probable background star kept in the sample}\\
{\footnotesize$^e$: emission in at least one Hydrogen line}\\
\end{longtable}


 \twocolumn

\section{Measurements of FHWM  and  estimates of projected rotational velocities for probable non-members }
\label{app:nomember} 

In this Appendix we list the main results 
 for the 15 stars probable non-members analyzed in this paper: GES ID, others names, the radial velocities measured from He and Si lines, Signal-to-noise ratio (S/N) of the observed spectra,  full width at half maximum (FWHM) measurements of the \ion{H}{i} lines 4388\AA\ and 4471\AA, The \Vsini\ estimates derived through the interpolation of the measured FWHM, and the average \Vsini\ values from \ion{H}{i} lines.

\begin{table*}
	\centering
	\caption{FHWM measurements and \Vsini\ estimates for 15 early-type stars in the Carina Nebula marked as probable non-member}
	\label{tab:nomember}
	\begin{tabular}{clccccccc} 
		\hline
   &   & V{r} (\kms)	&	&	\multicolumn{2}{c}{FWHM  (\AA)}	& \multicolumn{3}{c}{\Vsini\ (\kms)} 		\\
GES ID	& other ID	&  	& SNR	& 4388\AA & 4471\AA	&  4388\AA	& 4471\AA	&	Average	\\
	\hline
10442843-5934197$^{f}$	& 2MASS J10442842-5934196	&	 $-1\pm$1 	&124	&	2.85	&	3.48	&	108	&	113	&	110$\pm$3	\\
10441243-5934091$^{b}$	& 2MASS J10441242-5934091	&	 $-6\pm$2 	&191	&	6.30	&	6.69	&	260	&	258	&	259$\pm$1	\\
10441678-5920096$^{f}$	& CPD-58 2634			&	$-18\pm$1 	&176	&	3.02	&	3.42	&	109	&	101	&	105$\pm$6	\\
10441792-5925204$^{b}$ 	& 2MASS J10441791-5925204	&	$-13\pm$1 	&184	&	4.61	&	5.00	&	181	&	178	&	180$\pm$2	\\
10443090-5914461$^{b}$	& 2MASS J10443089-5914461     	&	 $-7\pm$3 	&27	&	--	&	2.61	&	--	&	85	&	85		\\
10444276-5921511$^{f}$	& ALS 15 857			&	  $+3\pm$2	&131	&	5.01	&	5.66	&	209	&	218	&	213$\pm$7	\\
10445888-5942231$^{b}$	& 2MASS J10445888-5942230	&	$-19\pm$2	&66	&	2.07	&	2.26	&	59	&	62	&	60$\pm$2	\\
10451063-5945126$^{f}$	& 2MASS J10451062-5945126	&	$-12\pm$2 	&40	&	--	&	1.29	& --		&	14	&	14		\\
10420950-6002267$^{b}$	& 2MASS J10420949-6002265	&	 $-3\pm$2 	&69	&	3.00	&	3.65	&	115	&	118	&	116$\pm$3	\\
10431946-5944488$^{f}$	& 2MASS J10431945-5944488	&	 $-6\pm$2	&27	&	--	&	1.44	&	--	&	32	&	32		\\
10442108-5956208$^{b}$	& 2MASS J10442107-5956207	&	 $-3\pm$2	&70	&	3.79	&	4.32	&	145     &	146	&	146$\pm$0	\\
10443239-5954458$^{f}$	& 2MASS J10443239-5954457	&	  $+5\pm$0	&28	&	--	&	7.20	&	--	&	281	&	281		\\
10424477-6005021$^{b}$	&[AHP2016] OBc 17	 	&	$-10\pm$7 	&131	&	5.27	&	5.87	&	222	&	231	&	226$\pm$7	\\
10454461-5950411$^{b}$	& ALS 16082			&	$-35\pm$1	&52	&	4.42	&	5.03	&	187	&	195	&	191$\pm$6	\\
10355661-5812407$^{b}$	&  CPD-57 3520    		&  	$-15\pm$1 	& 285   &       4.9	&       5.2	&    	197	&    	186	&    	191$\pm$8	\\
\hline
\hline
  \end{tabular}
  \\
{\footnotesize$^f$: foreground probable non-member}\\
{\footnotesize$^b$: background probable non-member}\\
  \end{table*}

\bsp	
\label{lastpage}
\end{document}